\documentclass[aps,pra,twocolumn,,superscriptaddress,floatfix]{revtex4-1}
\usepackage[svgnames]{xcolor}
\usepackage[colorlinks=true,urlcolor = purple,linkcolor=teal,citecolor=magenta,bookmarks=false]{hyperref}
\usepackage{amsfonts,amsmath,amssymb}
\usepackage{graphicx}
\usepackage{dsfont}

\usepackage{tikz}
\usetikzlibrary{arrows,topaths,shapes.geometric,decorations.markings,shadows,positioning,
plotmarks}
 
\usepackage{pgfplots}

\def\ii{{\rm i}}
\newcommand{\dd}{{\rm d}}
\def\bra#1{\mathinner{\langle{#1}|}}
\def\ket#1{\mathinner{|{#1}\rangle}}

\def\ol#1{\bar{#1}}

\def\ua{{\uparrow}}
\def\da{{\downarrow}}

\def\hstar{\,\hat{\star}\,}

\newcommand{\uu}{{\mathfrak{u}}}

\newcommand{\calC}{\mathcal{C}}
\newcommand{\calD}{\mathcal{D}}

\newcommand{\fs}{\mathfrak{s}}

\begin{document}

\title{Super-diffusion in one-dimensional quantum lattice models}

\author{Enej Ilievski}
\affiliation{Institute for Theoretical Physics Amsterdam and Delta Institute for Theoretical Physics,
University of Amsterdam, Science Park 904, 1098 XH Amsterdam, The Netherlands}

\author{Jacopo De Nardis}
\affiliation{D\'epartement de Physique, Ecole Normale Sup\'erieure,
PSL Research University, CNRS, 24 rue Lhomond, 75005 Paris, France}

\author{Marko Medenjak}
\affiliation{Faculty of Mathematics and Physics, University of Ljubljana, Jadranska 19, 1000 Ljubljana, Slovenia}

\author{Toma\v{z} Prosen}
\affiliation{Faculty of Mathematics and Physics, University of Ljubljana, Jadranska 19, 1000 Ljubljana, Slovenia}

\date{\today}

\begin{abstract}
We identify a class of one-dimensional spin and fermionic lattice models which display diverging spin and charge diffusion constants,
including several paradigmatic models of exactly solvable strongly correlated many-body dynamics such as the 
isotropic Heisenberg spin chains, the Fermi-Hubbard model, and the t-J model at the integrable point. Using the hydrodynamic transport 
theory, we derive an analytic lower bound on the spin and charge diffusion constants by calculating the curvature of the corresponding
Drude weights at half filling, and demonstrate that for certain lattice models with isotropic interactions some of
the Noether charges exhibit super-diffusive transport at finite temperature and half filling.
\end{abstract}


\maketitle


\paragraph*{Introduction.}

Understanding the microscopic mechanisms for the emergent macroscopic laws in many-body systems poses a fundamental question
in condensed matter physics. Despite a long tradition, the question has mostly been pursued by studying certain simple classical 
dynamical systems~\cite{Spohn1991}, such as elastically colliding rigid objects~\cite{Lebowitz64,LP67}, whereas much less is known 
about strongly-correlated quantum dynamics.

From the theoretical viewpoint, holographic theories \cite{Hartnoll2014,PhysRevLett.117.091601} and solvable systems in one dimension 
play an instrumental role in this context thanks to many powerful methods which enable explicit analytical calculations.
Exactly solvable models display anomalous transport behavior characterized by singular conductivities~\cite{PhysRevB.55.11029,1742-5468-2014-9-P09037,PhysRevLett.108.227206,PhysRevLett.82.1764,PhysRevB.84.155125,PhysRevB.95.060406,Prosen11,PhysRevLett.111.057203,
PhysRevB.87.245128,Ilievski2012,PhysRevB.91.115130,Mastropietro2018,1804.04476}. In contrast, very little is known about
the regular part of DC conductivities which characterize the sub-ballistic time scales, save for a few numerical studies typically suffering from strong finite-size or finite-time
effects~\cite{1367-2630-15-7-073010,PhysRevLett.112.120601,PhysRevB.89.075139,PhysRevB.90.155104,PhysRevB.88.195129,1702.08894}.
Exactly solvable interacting models are naturally tailored not only to tackle this problem in a rigorous manner,
but moreover permit efficient numerical
simulations~\cite{PhysRevB.96.245117,1367-2630-12-4-043001,1367-2630-19-3-033027,PhysRevB.93.205121,1742-5468-2016-6-063103,PhysRevB.92.205103,
PhysRevB.90.155104} and sometimes allow for experimental realizations~\cite{Hirobe2016,Hild14,Maeter13,Boll2016,1805.10990}.
Yet, even in a very simple interacting system, such as the integrable Heisenberg spin-$1/2$ chain,
the status of the spin dynamics on the sub-ballistic scales remains unresolved despite several recent numerical 
efforts; depending on the choice of parameters, the model shows a wide range of transport phenomena, ranging from 
ideal transport to diffusion and in some cases even super-diffusion
\cite{0295-5075-97-6-67001,Ljubotina_nature,Stephan17,Misguich17,CDV18,1711.11214}. 
It is thus reasonable to regard the Heisenberg spin-$1/2$ chain and other integrable models as exactly solvable representative models 
for various universality classes of transport behaviour exhibited by generic many-body quantum systems.

In this Letter, we report on a class of quantum spin and electron models which exhibit diverging spin and charge diffusion 
constants in thermal equilibrium with no charge or spin imbalance (i.e. at half-filling), despite the absence of ideal transport.
We build on an earlier proposal of ref.~\cite{PhysRevLett.119.080602} which relates
diffusion constants to the corresponding Drude weights in the vicinity of half-filled thermal states. Here we find a reinterpretation 
of the diffusion bound and optimize it in the framework of the hydrodynamic linear transport theory developed
in \cite{IN_Drude,SciPostPhys.3.6.039}.
We derive an analytic closed-form expression for the lower bound in the limit of infinite temperatures, and evaluate it for several 
paradigmatic interacting quantum lattice models. 

By explicitly calculating the bound on the diffusion constant, we show that for several models with \textit{isotropic interactions}, 
invariant under a continuous non-abelian (and possibly graded) Lie group $G$, that the conserved Noether charges
(e.g. spin or electron charges) belonging to the $SU(2)$ sector of the model exhibit \emph{super-diffusive} behavior in a half-filled 
state at any finite temperature. As prototypical examples we will focus on the Heisenberg spin chain and the Fermi--Hubbard chain.

\paragraph*{Summary.}

The central result of this work is an analytical lower bound on the spin/charge diffusion constants
for a family of interacting many-body one-dimensional lattice systems. Let $\hat{Q}=\sum_x \hat{q}_x$ denote a conserved
$U(1)$ charge of the model, with density $\hat{q}$
satisfying a local conservation law $\partial_{t}\hat{q}_{x}(t)+\partial_{x}\hat{j}_{x}(t)=0$.
The corresponding diffusion constant is defined via the Kubo formula
\begin{equation}\label{diffusionconst}
D^{(q)}(\beta) = \lim_{T\to \infty} \frac{\beta}{\chi^{(q)}(\beta)}\sum_{x}
\int_{0}^{T}\dd t \left\langle \hat{j}^{(q)}_{x}(t)\hat{j}^{(q)}_{0}(0) \right\rangle,
\end{equation}
where $\langle \bullet \rangle$ is the expectation value with respect to the grand-canonical Gibbs ensemble
$\hat{\varrho}_{\rm GC}(\beta)\simeq \exp{(-\beta \hat{H}+\sum_{i}2h_{i}\hat{N}_{i})}$ at inverse temperature $\beta$,
with $\hat{N}_{i}$ denoting the globally conserved $U(1)$ charges of the model including $\hat{Q}$,
$\chi^{(i)}(\beta)=\partial^{2}f(\beta)/\partial h^{2}_{i}$ denote
the static susceptibilities, and $f$ is the grand-canonical free energy\footnote{While in principle one should use the Kubo--Mori 
inner product, the latter reduces (under certain mild assumptions \cite{Ilievski2012} which do not affect our results) to the commonly 
used grand-canonical averaging.}.

We shall avoid a general formulation and rather concentrate on two prominent interacting systems which often play a 
pivotal role in the studies of strongly correlated one-dimensional materials, the anisotropic Heisenberg spin chain and
the Fermi--Hubbard model.
These exactly solvable systems feature stable interacting particle excitations which undergo a completely elastic scattering.
Consequently, the thermal average of the current density generally involves a dissipation-free component, implying
a singular DC conductivity characterized by a finite Drude weight
\begin{equation}
\calD^{(q)}(\beta) = \lim_{T\to \infty}\frac{\beta}{2T}
\sum_{x}\int_{0}^{T}\dd t \left\langle \hat{j}^{(q)}_{x}(t)\hat{j}^{(q)}_{0}(0) \right\rangle,
\end{equation}
which signals \emph{ballistic} transport. Drude weights can be efficiently computed within the
hydrodynamic approach developed in \cite{PhysRevLett.117.207201,PhysRevX.6.041065}, essentially exploiting
the fact that the net effect of inter-particle interactions (which are fully accounted for by a two-body scattering amplitude)
in the thermodynamic limit manifests itself as renormalization of particles' bare quantities in the presence of a
finite-density many-body background (e.g. a Gibbs thermal state or Generalized Gibbs states \cite{Pasquale-ed,1742-5468-2016-6-064007,1742-5468-2016-6-064002,PhysRevLett.115.157201,IQC17}), commonly referred
to as \emph{dressing} (see e.g
\cite{SciPostPhys.2.2.014,PhysRevB.96.115124,PhysRevLett.119.220604,PhysRevLett.119.195301,PhysRevB.96.081118}).
Spectra of solvable models are parametrized in terms of particle excitations. We label them by a discrete index $A$ counting over
(typically infinitely many) particle types, and a continuous rapidity variable $u$ encoding their bare momenta
$k_{A}(u)$ and energies $e_{A}(u)$. The dressing of a bare quantity $q_{A}$ is expressible as a linear transformation
$q_{A}\mapsto q^{\rm dr}_{A}$, while the effective velocities of propagation are obtained from the dressed dispersion relations, 
$v^{\rm eff}_{A} = \partial \varepsilon_{A}/\partial p_{A} = \varepsilon^{\prime}_{A}/p^{\prime}_{A}$, where
$p^{\prime}_{A} = (k^{\prime}_{A})^{\rm dr}$ and $\varepsilon_{A} = (e^{\prime}_{A})^{\rm dr}$, with
prime denoting the rapidity derivative.
In this picture, the hydrodynamic mode decomposition of the Drude weight reads \cite{SciPostPhys.3.6.039,PhysRevB.96.081118}
\begin{equation}
\label{eq:drude_decomposition}
{\calD}^{(q)}(\beta) =
\frac{\beta}{2}\sum_{A}\int \dd u\,\mathcal{D}_{A}(u)\left[q^{\rm dr}_{A}(u)\right]^{2},
\end{equation}
where $\mathcal{D}_{A}(u)= \rho_A(u) (1- \vartheta_{A}(u)) [v^{\text{eff}}_{A}(u)]^2$ is the `Drude kernel'
and  $q^{\rm dr}_{A}(u)$ are the dressed charges of individual excitations with respect to an equilibrium state
(defined in \cite{SM}). Dependence on the reference equilibrium state enters through the rapidity distributions
$\rho_{A}(u)$, which are uniquely determined by the mode occupation (filling) functions
$\vartheta_{A}(u) = \rho_{A}(u)/[2\pi\,\sigma_{A}p^{\prime}_{A}(u)]$ ($\sigma_{A}={\rm sgn}(k^{\prime}_{A}(u))$).

\paragraph*{Lower bound on diffusion.}
In the half-filled equilibrium states, the spin/charge Drude weight vanishes due to the symmetry reasons despite integrability.
To characterize transport on sub-ballistic time-scales we exploit a useful relation between the diffusion constant and the
curvature of the Drude weight with respect to the filling parameter, proposed in~\cite{PhysRevLett.119.080602}. Consequentially,
the relation provides a non-vanishing lower bound on diffusion provided the Drude weight vanishes at most quadratically as a function
of the filling parameter. This condition is satisfied for the half-filled thermal states in particle-hole symmetric lattice models 
considered in this work.

To briefly outline the idea of the lower bound, we imagine a small gradient of the charge density imposed across the system and 
subsequently measure the induced current. The current (initially localized at the origin) spreads only over a finite portion of the 
system in a finite amount of time due to the Lieb--Robinson causality. This means that at finite times on 
the relevant sublattice, the probability of measuring the current in the sector away from half-filling is non-zero. In these sectors the current grows indefinitely with time, however
the probability of the system being away from half-filling vanishes with the system size. The interplay of
vanishing probability and diverging conductivity permits to obtain a lower bound on the diffusion 
constant, reading~\cite{PhysRevLett.119.080602}
\begin{equation}
D^{(q)}(\beta)\geq \frac{1}{8 \beta \chi^2(\beta) v_{LR}}\partial_h^2 \mathcal{D}^{(q)}(\beta,h)\Big|_{h=0},
\end{equation}
where $ v_{LR} = \text{max}_{u,A} v^{\text{eff}}_A(u)$ is the Lieb-Robinson velocity. 
In particular, in the high-temperature limit the bound becomes
\begin{equation}\label{bound1}
D^{(q)}(0)\geq \lim_{\beta \to 0} \frac{18}{\beta (d^2-1)^2 \, v_{LR}}\partial_{h}^2 \mathcal{D}^{(q)}(\beta,h)\Big|_{h=0},
\end{equation}
where we assumed that the local degrees of freedom carry charge $q\in \{-\tfrac{1}{2}(d-1),...,\tfrac{1}{2}(d-1)\}$.

\paragraph*{Solving the dressing equations.}
The Drude weight and its curvature can be expressed in terms of dressed quantities \eqref{eq:drude_decomposition}.
Given the full set of equilibrium occupation functions $\vartheta_{A}$, the dressing equations take the form of
coupled linear integral equations, cf. \cite{SM}. Functions $\vartheta_{A}$ are determined by minimizing the free energy as 
a functional of the densities $\rho_{A}$. This requires to solve a system of non-linear integral equations, which is only possible
numerically using an iteration scheme, except in two extreme cases corresponding to either the ground states or
the high-temperature limit. In the latter case, the occupation functions become momentum-independent and the
dressing transformation becomes an algebraic system.
For the class of rotationally symmetric solvable spin and fermion lattice Hamiltonians considered here,
the dressing equations admit an analytic group-theoretic solution, as explained in detail in \cite{SM}.
This permits us to obtain a closed-form expression for the bound \eqref{bound1} (when it is finite), and
rigorously establish the occurrence of super-diffusion signalled by a divergent bound. Importantly, since the divergence is
a result of a particular dependence on the dressed properties of particles with large bare spin/charge,
the main statement about the super-diffusive dynamics remains valid even at finite temperatures.
We note that in our calculations we take into account the exact dressed dispersion relations of interacting excitations,
and our results cannot be accessed with alternative approaches, such as effective field-theoretical 
methods~\cite{PhysRevLett.103.216602,PhysRevB.83.035115,1367-2630-17-10-103003,AKT06} or
semi-classical approximations~\cite{SD97} which fail to capture the essential contributions of the bound states.

\begin{figure}[htb]
\centering
\includegraphics[width=0.75\hsize]{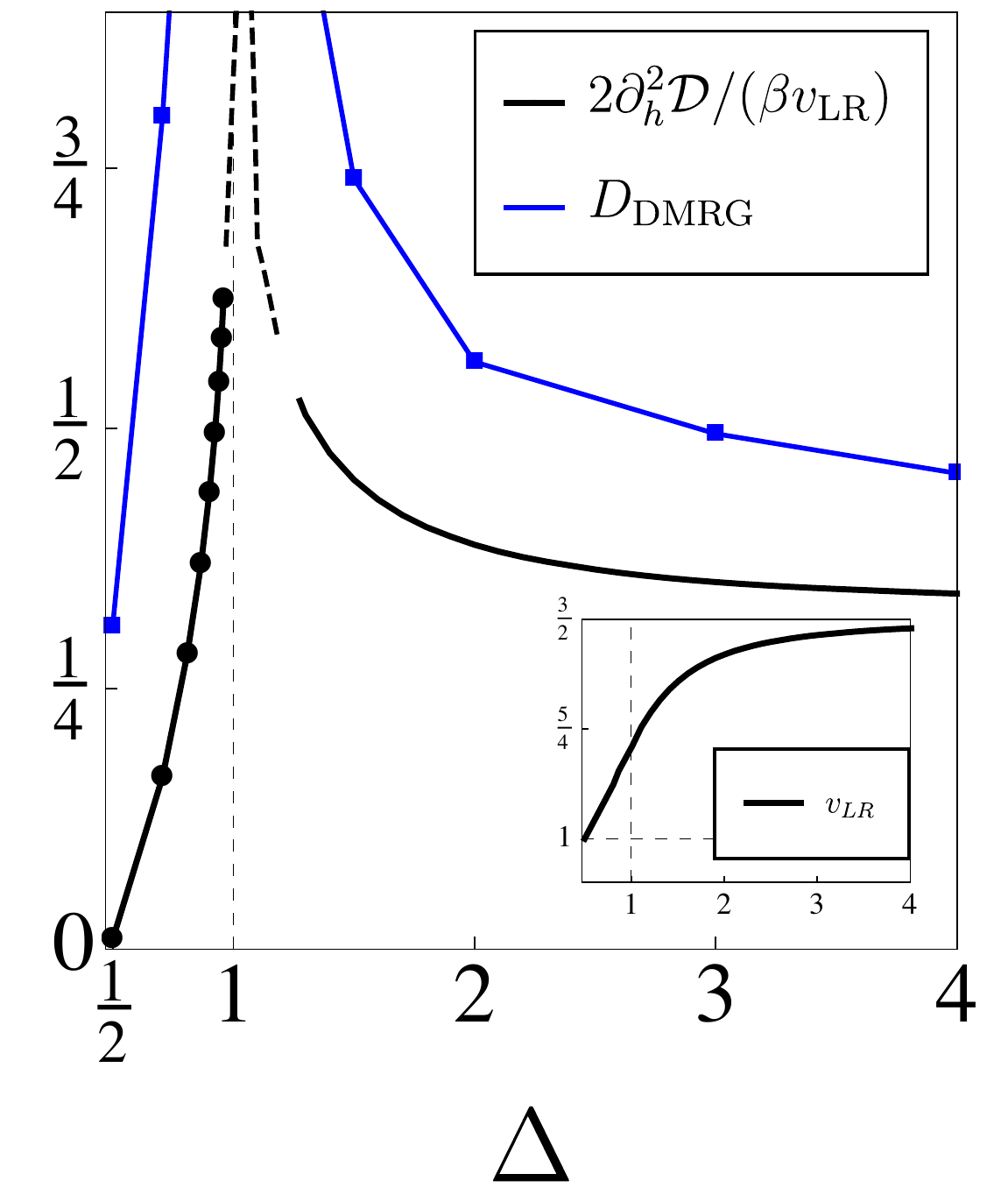}
\caption{XXZ chain at infinite temperature: the black curve shows diffusion bound \eqref{bound1},
$D^{(m)}\geq 2\calC^{(m)}(0)/(\beta v_{LR})$, and the blue points display numerical values of the spin diffusion constant 
obtained by tDMRG in refs.~\cite{PhysRevB.89.075139} and \cite{1367-2630-19-3-033027}. The logarithmic divergence close
to $\Delta=1$ is indicated by the dashed line.  Notice that the bound does not vanish even in the Ising limit
$\Delta \to \infty$, in contrast to the dissipative case \cite{PhysRevLett.106.220601}. Inside the gapless interval
we display $\Delta=\cos{\pi/\ell}$ with $\ell = 3,\ldots,10$.}
\label{fig}
\end{figure}

We subsequently concentrate on the transport of global $U(1)$ charges, such as the total magnetization
$\hat{S}^{z}\equiv \sum_j \hat{S}^z_j$, and/or the total electron charge
$\hat{N}_{\rm e}\equiv \frac{1}{2}\sum_{j,\sigma=\uparrow,\downarrow} \hat{c}^{\dagger}_{j,\sigma} \hat{c}_{j,\sigma}$.
We consider the half-filled spin/charge sectors, where the Drude weights vanishes as
\begin{equation}
\calD^{(q)}(\beta,h) = \calC^{(q)}(\beta) \frac{h^{2}}{2} + \ldots \qquad {\rm for}\quad h\sim 0,
\end{equation}
and evaluate the bound \eqref{bound1}. The Drude weight curvature reads
\begin{equation}
\label{eqn:bound_decomposition}
\calC^{(q)}(\beta) = \frac{\beta}{2}\sum_{A}\int \dd u\,\mathcal{D}_{A}(u)\,
\partial^2_{h}\left[q^{\rm dr}_{A}(u)\right]^{2} \Big|_{h=0}.
\end{equation}

In exactly solvable interacting quantum lattice models the elementary excitations which carry spin and charge typically
form bound states. Let integer $s$ denote their `bare charge' (or `bare mass'), i.e. the number of constituents within a bound state; 
for instance, in a spin system, such as the Heisenberg spin chain, $s$ pertains to the number of bound magnons in multi-magnon 
excitations, while in an electron system (e.g. the Fermi--Hubbard model) $s$ can be 
the number of bound spin-full electrons which form spin singlet states etc. 
Moreover, if the Hamiltonian has a global rotational symmetry of a (graded) Lie group $G=SU(N|M)$
(with scattering amplitudes being rational functions of the scattering momenta), the number of distinct bound states
is \emph{infinite}, i.e. $s$ can be arbitrarily large.
We found that, for such models the Drude weight curvature per particle decreases as $\sim 1/s$
for large $s$, yielding a (logarithmically) divergent diffusion lower bound after summing over all the particle types.

\begin{figure}[h]
\centering
\includegraphics[width=0.75\hsize]{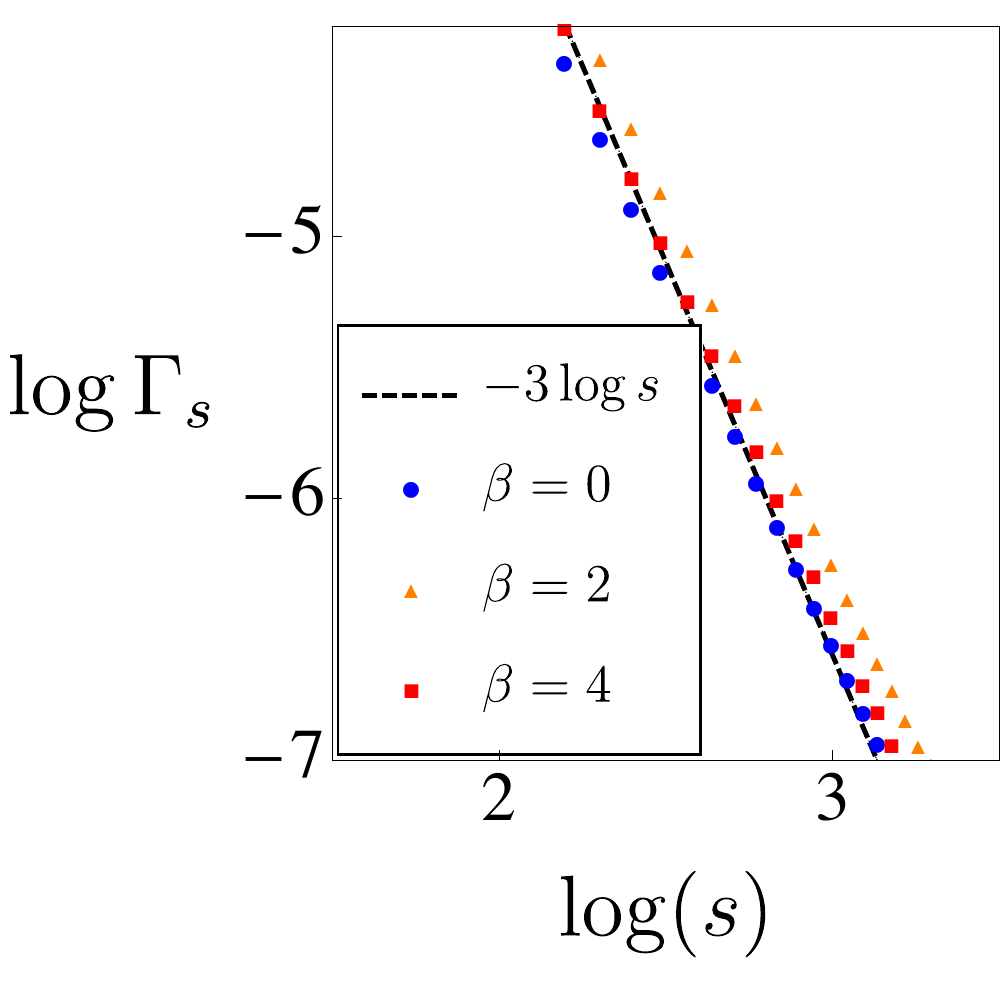}
\caption{Large-$s$ scaling of
$\log \Gamma_s = \log  \int_{-\infty}^{\infty}\dd u p^{\prime}_{s}(u)\left[v^{\rm eff}_{s}(u)\right]^{2}$,
confirming the asymptotic of Eq.~\eqref{eq:v_scaling} for the isotropic Heisenberg chain for various temperatures, showing 
that the large-$s$ scaling is independent of $\beta$.}
\label{fig2}
\end{figure}

\paragraph*{Anisotropic Heisenberg spin-1/2 chain.}
The simplest model which features several distinct transport regimes is the Heisenberg XXZ spin-$1/2$ chain,
\begin{equation}
\hat{H}_{\rm XXZ} = \sum_{j=1}^{L}\left(\hat{S}^{x}_{j}\hat{S}^{x}_{j+1} + \hat{S}^{y}_{j}\hat{S}^{y}_{j+1}+
\Delta \hat{S}^{z}_{j}\hat{S}^{z}_{j+1}\right),
\label{eqn:XXZ_Hamiltonian}
\end{equation}
with gapless (gapped) spectrum for $|\Delta|\leq 1$ ($|\Delta|>1$).
The interaction anisotropy has a profound influence on the spin transport, shortly summarized below.
The hydrodynamic representation of the spin Drude weight curvature reads
\begin{equation}
\calC^{(m)}(\beta)  = \frac{\beta}{2}\sum_{s\geq1} \int \frac{\dd u}{2 \pi} \vartheta_{s}(1-\vartheta_{s}) p^{\prime}_{s}
\left[v^{\rm eff}_{s}\right]^{2}\frac{\partial^2 \left[ m^{\rm dr}_{s}\right]^{2}}{\partial h^2} \Big|_{h=0},
\label{eq:boundXXZ}
\end{equation}
where $m^{\rm dr}_{s}=\partial_{2h}\log (\vartheta^{-1}_{s}-1)$.
Our conclusions are:\\
Exactly at the $SU(2)$ isotropic point $\Delta=1$, the finite-temperature spin diffusion constant $D^{(m)}$
\emph{diverges} in the limit of half filling $h \to 0$. This can be inferred from the large-$s$ scaling of the dressed
spin (magnetization), mode occupation functions and dressed dispersion relations,
\begin{align}  
\label{eq:effective_spin}
&  m^{\text{dr}}_{s}(h) \simeq \tfrac{1}{3}\left(s+\kappa(\beta)\right)^{2} h  + \mathcal{O}(h^{3}),\\
\label{eq:occupation_scaling}
&  \lim_{h\to 0}\vartheta_{s}(h) \simeq  {\left(s+\kappa(\beta)\right)^{-2}},\\
\label{eq:v_scaling}
& \lim_{h\to 0}\int_{-\infty}^{\infty}\dd u\, p^{\prime}_{s}(u)\left[v^{\rm eff}_{s}(u)\right]^{2} \simeq \frac{1}{s^3},
\end{align}
for some (unknown) temperature-dependent function $\kappa(\beta)$. The above large-$s$ asymptotics holds for any finite value of 
$\beta$. The finite-temperature behaviour \eqref{eq:v_scaling} is also confirmed numerically, see Fig.\ref{fig2}.
In the $\beta \to 0$ limit however, relations \eqref{eq:effective_spin} and \eqref{eq:occupation_scaling} indeed become 
equalities valid for all values of $s\geq 1$, with $\kappa(0)=1$.

In the gapped regime, anisotropy $\Delta= \cosh \eta$ breaks the $SU(2)$ symmetry of the interaction to $U(1)$.
In the limit of infinite temperature and vanishing chemical 
potential, the dressed spin and mode occupations functions of bound magnons remain the same as in the isotropic case, cf. 
Eqs.~\eqref{eq:effective_spin},\eqref{eq:occupation_scaling}.
Notice that $\vartheta_{s}$ and $m^{\rm dr}_{s}$ become independent of $u$ for large $s$.
The key difference now is that the bare dispersion of bound magnons become $\eta$-dependent functions. In particular,
the rapidity-dependent part of Eq.~\eqref{eq:boundXXZ} scales as
\begin{equation}
\int_{-\pi/2}^{\pi/2} \frac{\dd u}{2 \pi}p^{\prime}_{s}(u)\left[v^{\rm eff}_{s}(u)\right]^{2} \simeq e^{-\eta s},
\end{equation}
i.e. is exponentially suppressed for large bound states. Contrary to the isotropic case, exponential convergence in $s$
results in a finite spin diffusion lower bound \eqref{bound1}.

The gapless regime $|\Delta|<1$ is rather exceptional, with a positive finite-temperature spin Drude weight
even in the half-filled sector \cite{PhysRevLett.82.1764,Ilievski2012,IN_Drude}, with a non-continuous dependence on $\Delta$.
Still, it is interesting to ask whether the sub-ballistic corrections to spin transport are normal, diffusive or anomalous
sub/super-diffusive. The thermodynamic particle content of the model in this regime is quite involved
(see \cite{Takahashi1999}) and, in distinction to the gapped regime, changes depending on the value
of $\Delta$ \cite{PhysRevB.96.020403,IN_Drude}. For simplicity we restrict ourselves to discrete points
$\Delta= \cos \pi /\ell$, for integer $\ell \geq 3$ (the Drude curvature at $\Delta=0$ is not positive, consistently
with the vanishing diffusion constant at the free fermionic point \cite{Spohn_JMP}),
where the spectrum consist of $\ell$ distinct particle types. For $s=1,\ldots,\ell-2$, the particles represent bound states of $s$ 
magnons whose high-temperature dressed spin is given by Eq.~\eqref{eq:effective_spin} which therefore vanishes as $h\to 0$.
There is an extra (exceptional) doublet of particles carrying finite dressed spins
$m^{\rm dr}_{A}=\ell/2 \pm \kappa_{\ell} h $ ($\kappa_{\ell}>0$), for $A=\ell-1,\ell$, charged under the non-unitary local 
conservation laws found in \cite{Prosen11,PhysRevLett.111.057203}, which are responsible for the non-vanishing of spin Drude weight
even at half filling~\cite{IN_Drude}.
A finite contribution to the curvature $\calC^{(m)}$ is obtained by subtracting a finite Drude weight
$\mathcal{D}^{(m)}= \sum_{A=\ell-1,\ell} \int \dd u \rho_{A}(u)(1-\vartheta_{A}(u)) (v^{\text{eff}}_{A}(u) \ell/2)^{2}$
and expanding the remainder to the second order in $h$. We find a finite lower bound for all $\ell<\infty$ which diverges
as $\ell \to \infty$, namely $\Delta \to 1^{-}$, as shown in Fig.~\ref{fig}.~\footnote{At discrete points
$\Delta = \cos  \frac{\pi}{\ell + 1/\nu}$, the number of magnonic bound states is $\ell +\nu$ and, therefore, after subtracting a 
finite Drude weight, the Drude curvature diverges as $\nu \to \infty$, similarly to the case
$\Delta \to 1^{-}$. This suggests that the spin diffusion constant, similarly to the Drude weight, is not a smooth function of 
$\Delta$, and that it diverges almost everywhere for $\Delta \in [-1,1]$.}

\paragraph*{Fermi--Hubbard model.}
Another class of models of particular importance are lattice models of fermions, the most prominent example
being the one-dimensional Fermi--Hubbard model describing spin-full electrons interacting via Coulomb repulsion,
\begin{equation}
\hat{H}_{\rm H} = -\sum_{j=1}^{L} \sum_{\sigma \in \ua,\da}\hat{c}^{\dagger}_{j,\sigma}\hat{c}_{j+1,\sigma}+
\hat{c}^{\dagger}_{j+1,\sigma}\hat{c}_{j,\sigma} + 4\uu\sum_{j=1}^{L}\hat{V}^{\rm H}_{j},
\label{eqn:Hubbard_Hamiltonian}
\end{equation}
with $\hat{V}^{\rm H}_{j} = \sum_{j=1}^{L}\left(\hat{n}_{j,\ua}-\tfrac{1}{2})(\hat{n}_{j,\da}-\tfrac{1}{2}\right)$.
The spin and charge excitations both participate in the formation of bound states.
The particle content consists of individual spin-up electrons, spin-singlet compounds of $2a$ electrons with
($a\in \mathbb{N}$) and charge-less bound states of $s$ spin excitations with bare spin ($s\in \mathbb{N}$).
Although spin and charge degrees of freedom mutually interact and undergo a non-trivial dressing,
the transport of both spin and charge are in \emph{qualitative} agreement with the isotropic Heisenberg chain:
in the vicinity of the half-filled regime $h\to 0$ where $\calD^{(m)}(\beta)$ vanishes, the dressed spin and thermal 
occupation functions scale with $s$ as $m^{\rm dr}_{s}(h)\sim h\,s^{2}$ and $\lim_{h\to 0}\vartheta_{s}(h)\sim s^{-2}$, respectively,
with no dependence on charge chemical potential $\mu$ associated to the conservation of the number of electrons.
An analogous reasoning applies for the transport of electron charge, see \cite{SM} for further details.
Numerical evaluation shows that the momentum-dependent part of $\calD_{A}$ for the spin-carrying bound states once again scales as
in Eq.~\eqref{eq:occupation_scaling}, implying a (logarithmically in $s$) diverging spin diffusion bound \eqref{bound1}.

\paragraph*{Higher spins and higher rank symmetries.}
We have additionally solved the dressing equations for a family of integrable spin-$S$ isotropic Heisenberg chains, and for the
higher-rank $SU(N)$-symmetric lattice models which comprise $N-1$ species of interacting excitations.
The picture, exemplified above for the isotropic $S=1/2$ Heisenberg model ($N=2$), is qualitatively unchanged.
By virtue of $SU(N)$ invariance however, the statement now holds for all the components of the Noether charges.
Explicit results are reported in the Supplemental Material \cite{SM}, which includes
refs.~\cite{Zamolodchikov91,EKS92,Frahm99,GK11,FQ12,Volin12,String_charge,Kazakov16}.

\paragraph*{Other fermionic models.}
We have also inspected the $SU(2|2)$-symmetric (EKS model \cite{EKS92}) and $SU(2|1)$-symmetric (t-J model) fermionic
lattice models of spin-carrying electrons, where the conclusions do not change provided the conserved $U(1)$ charge $\hat{Q}$
belongs to a bosonic (i.e. even) $SU(2)$ sector.
Notice however that, in addition to the conserved total magnetization $\hat{S}^{z}$, the $SU(2|1)$-invariant integrable t-J model 
conserves the total electron charge $\hat{N}_{e}$ which (in distinction to the total spin and charge in the
Fermi--Hubbard model) corresponds to a global $U(1)$ charge which does not belong to the $SU(2)$ sector of the
full $SU(2|1)$ symmetry of the Hamiltonian. The absence of particle-hole symmetry for the electron charge implies
a finite charge Drude weight in any equilibrium state with a finite density of electrons,
and the diffusion bound cannot be employed there.
Likewise, in the $SU(2|2)$-symmetric model of spin-full electrons there exists, besides two 
independent spin $SU(2)$ sectors as in the Hubbard model, the third global $U(1)$ conserved charge (the Hubbard interaction 
$\hat{V}_{\rm H}$). The latter also yields a finite Drude weight for all values of chemical potentials (cf. \cite{SM} for
additional information).

\paragraph*{Conclusion.}

We identified and discussed a class of exactly solvable quantum lattice models with isotropic interactions where
Noether charges exhibit sub-ballistic transport with divergent diffusion constants. Super-diffusive transport is attributed to the 
existence of infinitely many bound states of magnons or electrons which behave at any finite temperature
(cf. Fig.~\ref{fig2} and Eqs.~\eqref{eq:occupation_scaling}, \eqref{eq:effective_spin}) as effective paramagnetic compounds
of spins (or electrons): their dressed spin (or charge) grows as $\sim h\,s^{2}$ with their bare mass $s$ for small
values of chemical potential $h$, and whose velocities decay proportionally to $1/s$.
We wish to stress that an infinite number of bound states in the spectrum is not a sufficient for a divergent diffusion constant, as 
shown explicitly in the gapped regime of the anisotropic Heisenberg spin-$1/2$ chain where, indeed, bound states acquire dressed 
velocities which get exponentially suppressed with their size.

There are several related aspects to be addressed in future. At the moment it is difficult to estimate the importance of
integrability for the observed anomalous behavior. Although we have excluded normal diffusion at half filling,
invoking only a lower bound precludes determining the exact super-diffusive dynamical exponent.
Indeed, numerical simulations on the isotropic quantum~\cite{Ljubotina_nature} and classical Heisenberg magnet~\cite{Bojan} give
a firm indication of dynamical exponent $z=2/3$ -- which is consistent with the Kardar-Parisi-Zhang  (KPZ)
universality \cite{PhysRevLett.56.889} (also observed in random unitary circuits in (1+1)D \cite{PhysRevX.8.021014}),
in contrast to the standard diffusive exponent $ z={1/2}$ observed in anisotropic models and strongly dissipative
XXZ chains \cite{SciPostPhys.3.5.033}.
The hope is that the models exhibiting super-diffusion identified in this paper can be viewed as representative models for
a broad super-diffusive universality class of quantum systems, possibly of the KPZ-type, whose precise determination
remains an open problem.

\textit{Acknowledgments.}
The authors thank C. Karrasch for providing tDMRG data, and thank F. Heidrich-Meisner for valuable comments.
J.D.N. acknowledges support from LabEx ENS-ICFP:ANR-10-LABX-0010/ANR-10-IDEX-0001-02 PSL*.
E.I. is supported by VENI grant number 680-47-454 by the Netherlands Organisation for Scientific Research (NWO). M.M. and T.P. 
acknowledge the support from the ERC Advanced grant 694544 – OMNES and the grant P1-0044 of Slovenian Research Agency.

\bibliography{Diffusion_bound}

\appendix
\onecolumngrid

\begin{center}
\textbf{{\Large Supplemental Material}}
\end{center}
\begin{center}
\textbf{{\large Super-diffusion in one-dimensional quantum lattice models}}
\end{center}

This Supplemental Material includes:
\begin{enumerate}
\item An exposition of the (nested) Thermodynamic Bethe Ansatz dressing formalism for a family of
integrable quantum lattice models comprising spin or electron degrees.
\item A short derivation of the lower bound on the charge diffusion constants from the curvature of the Drude weight.
\end{enumerate}

\section{Dressing formalism: Nested Bethe Ansatz}
The notion of \emph{dressing} is the central ingredient of the Bethe Ansatz formalism.
Integrable systems are interacting theories which host stable particle-like excitations which scatter elastically,
with no particle decay or production. Particles here refer to soliton-like objects which preserve their nature upon collisions.
While non-interacting particles, subjected to periodic boundary conditions, are describe by plane waves whose momentum $k$ obeys
the quantization  constraint $\exp{(\ii k\,L)}=1$ (irrespectively of other particles in the system),
the quantization rule for interacting particles on the other hand acquires an extra multiplicative phase factor,
\begin{equation}
e^{\ii k_{i}L}e^{\ii \Phi_{i}}=1 \qquad \Longleftrightarrow \qquad
e^{\ii p_{i}L} = 1,
\label{eqn:dressing}
\end{equation}
due to an accumulated two-body \emph{phase shift} $\delta_{2}$,
\begin{equation}
\Phi_{i} \equiv \sum_{i \neq j = 1}^{M}\delta_{2}(k_{i},k_{j}).
\end{equation}
The absence of diffraction means that multi-particle processes are completely factorizable in terms two-body collisions
described by the scattering phase $\delta_{2}$.
The dressing can be thought of as a renormalization of particles' bare momenta $k_{i}$, corresponding to
absorbing the scattering shift into a redefinition of the excitation momentum, $k_{i}\mapsto k^{\rm dr}_{i}\equiv p_{i}$.
In this case, $p_{i}$ depend implicitly on the dressed momenta of all other $M-1$ excitations present in a given eigenstate.

By virtue of integrability, a many-body scattering process is elastic (i.e. free of diffraction) and completely factorizes into
a sequence of two-body scattering events, irrespective of the ordering of the two-particle collisions. As a corollary, the set of 
outgoing momenta are simply a permutation of the momenta of incoming particles.\\

Computing the dressing for a state with finitely many excitations in a finite volume amounts to solve the celebrated Bethe Ansatz 
equations. In the thermodynamic limit -- defined by as the scaling limit $L,N\to \infty$ with $N/L$ kept fixed -- the
momenta can take continuous values, and equations \eqref{eqn:dressing} can be converted to a set of coupled
integral equation for the (dressed) momentum density.\\

\subsection*{Integrable graded lattice models}
We devote our analysis to the class of homogeneous one-dimensional integrable quantum lattice models with
$SU(N|M)$-symmetric Hamiltonians. Unless stated otherwise, the physical degrees of freedom associated to the lattice sites
belong to the fundamental representation of $\mathfrak{su}(M|N)$ graded Lie algebras. A convenient property of this family of models 
is that they can be treated in a uniform way, which can be attributed to the fact that they all share the same elementary scattering 
amplitudes
\begin{equation}
S_{j}(u) = \frac{u-j\tfrac{\ii}{2}}{u+j\tfrac{\ii}{2}},
\end{equation}
for $j\in \mathbb{N}$, describing interactions between various particle excitations in their spectra.
Specifically, the scattering amplitude associated to two elementary excitations,
denoted by $S_{1,1}(u,w)=\exp{(\ii \delta_{2}(u,w))}$, is a rational function of the particles' rapidity variables,
denoted by $u$ and $w$. The bare momentum $k_{j}$ of the \emph{fundamental} particle reads
\begin{equation}
k_{j}=k(u_{j}) = \ii \log S_{1}(u_{j}).
\end{equation}

Elementary particle excitations in addition participate in the formation of composite (bound) states.
The entire set of scattering amplitudes describing interactions among them is simply obtained by `fusing' the 
elementary scattering amplitudes,
\begin{equation}
S_{j,k}(u,w) = S_{|j-k|}(u,w)S_{j+k}(u,w)\prod_{m=1}^{{\rm min}(j,k)-1}S^{2}_{|j-k|+2m}(u,w).
\label{eqn:fused_scattering_amplitudes}
\end{equation}

We shall primarily be interested in interacting models described by $SU(N|M)$-symmetric Hamiltonians
\begin{equation}
\hat{H}^{N|M}=\sum_{j}\hat{h}^{N|M}_{j,j+1},
\label{eqn:graded_Hamiltonians}
\end{equation}
with interaction density $\hat{h}^{N|M}_{j,j+1}$ acting on adjacent lattice sites $j$ and $j+1$. Each lattice site is
associated the fundamental degree of freedom, that is the fundamental representation $V_{\square}\cong \mathbb{C}^{N+M}$
of $\mathfrak{su}(M|N)$. The latter formally a graded vector space spanned by vectors $v_{i}$ ($i=1\sim N+M$) equipped
with Grassmann $\mathbb{Z}_{2}$-parity $|i|\in \{0,1\}$.
There are $N$ bosonic states with the assigned parity $|i|=0$, and $M$ fermionic states with parity $|i|=1$.
The linear algebra of operators acting in the fundamental representation $V_{\square}$ is spanned
by the $\mathfrak{su}(N|M)$ generators $\hat{E}^{ij}$ which obey the following graded commutations relations
\begin{equation}
\left[\hat{E}^{ij},\hat{E}^{kl}\right] = \delta_{jk}\hat{E}^{il} - (-1)^{(|i|+|j|)(|k|+|l|)}\delta_{il}\hat{E}^{kj}.
\end{equation}

The grading can be assigned arbitrarily. Inequivalent gradings are in one-to-one correspondence with Kac--Dynkin diagrams, consisting
of $N+M-1$ nodes which are either bosonic when states $v_{i}$ and $v_{i+1}$ are of the same parity (open circles) or fermionic
when $v_{i}$ and $v_{i+1}$ have different parities (crossed circles).

The Hamiltonians \eqref{eqn:graded_Hamiltonians} on $V^{\otimes L}_{\square}$, with the interaction densities
$\hat{h}^{N|M}$ of the form
\begin{equation}
\hat{h}^{N|M} = 1-\hat{P}^{N|M},
\end{equation}
where
\begin{equation}
\hat{P}^{N|M}=(-1)^{|a||b|}\sum_{a,b=1}^{N+M}\hat{E}^{ab}\otimes \hat{E}^{ba},
\label{eqn:graded_permutation}
\end{equation}
denotes the \emph{graded} permutation operator on $V^{\otimes 2}_{\square}$.

The Hamiltonians $\hat{H}^{N|M}$ can be diagonalized by means of the \emph{nested Bethe Ansatz}. 
Below we summarize the main ingredients of this procedure. Notice that the full algebraic construction of
exact finite-volume eigenstates is not of our main concern. Instead, we shall only interested in the complete spectrum of
particle-like excitations and their properties.

\subsection*{Particle content}

The fundamental spin chain invariant under the $SU(N|M)$ group possesses $N+M-1$ types of elementary excitations.
The later can be associated with nodes of the corresponding Kac--Dynkin diagram, in a one-to-one fashion.
By fixing the grading, each highest-weight Bethe eigenstate is characterized by a unique set of rapidity variables
which come in of $N+M-1$ flavours,
\begin{equation}
\left\{u^{(k)}_{j}\big|j=1\sim N_{k};k=1\sim N+M-1\right\}.
\end{equation}
satisfying a coupled set of algebraic equations known as the \emph{nested Bethe Ansatz} equations
\begin{equation}
\exp{\Big(\ii p \big(u^{(\ell)}_{i}\big)L\Big)}
\prod_{k=1}^{N+M-1}\prod_{j=1}^{N_{k}}S_{\ell k}\left(u^{(\ell)}_{i},u^{(k)}_{j}\right)=-1,
\label{eqn:Bethe_equations_scattering}
\end{equation}
where $N_{k}$ is the number of Bethe roots of type $k$ associated to the $k$th Dynkin node.
The rational scattering amplitudes read explicitly,
\begin{equation}
S_{\ell k}\left(u^{(\ell)}_{i},u^{(k)}_{j}\right) =
\frac{u^{(\ell)}_{i}-u^{(k)}_{j}-\tfrac{\ii}{2}\mathcal{K}_{\ell k}}
{u^{(\ell)}_{i}-u^{(k)}_{j}+\tfrac{\ii}{2}\mathcal{K}_{\ell k}},
\label{eqn:scattering_amplitudes}
\end{equation}
and are parametrized with aid of the \emph{graded Cartan matrix},
\begin{equation}
\mathcal{K}_{\ell,k}=\delta_{\ell,k}\big((-1)^{|\ell|}+(-1)^{|\ell+1|}\big)-
(-1)^{|\ell+1|}\delta_{\ell+1,k}-(-1)^{|\ell|}\delta_{\ell-1,k},
\end{equation}
The latter depends on the choice of grading which, in particular, fixed the simple positive roots. Physically speaking,
this means to select a particular reference (Bethe) vacuum state.
\\

\paragraph*{Particles and rectangular partitions.}
Elementary excitations can be either of bosonic or fermionic type.
To describe the complete thermodynamic spectra of graded spin chains, one has to understand the formation of bound
particles, being certain composites of the elementary excitations. Such compounds are known in the literature the `Bethe strings'.
For the class of $SU(N|M)$-symmetric homogeneous lattice models considered here (cf. Eq.~\eqref{eqn:graded_Hamiltonians}),
the complete particle spectrum turns out to bijectively correspond to the finite-dimensional (unitary) 
irreducible representations associated to \emph{rectangular partitions}, that is Young tableaux $[a,s]$ with $a$ rows and $s$ columns.
A distinguished property of rectangular irreducible representations is that they constitute a closed set of fusion rules.
This is in fact dictated by the composition rule of the underlying \emph{classical} Lie algebra $\mathfrak{g}$. Accordingly,
it is most natural label the particle species (bound states included) by a pair of positive integers $A=(a,s)$.
In addition, each particle is characterized by a continuous rapidity variable $u$.\\

\paragraph*{Inter-particle interactions.}
The entire class of graded lattice models with $\mathfrak{g}=\mathfrak{su}(N|M)$ symmetry shares a common `tight-binding'
four-vertex incidence (adjacency) matrix
\begin{equation}
I_{AB} \equiv I_{(a,s),(a^{\prime},s^{\prime})} =
\delta_{a,a^{\prime}}(\delta_{s,s-1}+\delta_{s,s+1}) + \delta_{s,s^{\prime}}(\delta_{a,a-1}+\delta_{a,a+1}),
\label{eqn:indicende_matrix_abstract}
\end{equation}
which compactly encodes the fusion rules of the scattering amplitudes and can be used to express effective interactions 
among the particles. Its form reflects the internal structure of the bound states of elementary excitations.
\\

The boundary conditions for $I_{AB}$ depend on the rank of the algebra $\mathfrak{g}$ and the number of bosonic states $N$.
This is best understood by recalling the bijective correspondence between the particles and rectangular partitions,
which tells that the particles nicely arrange on a two-dimensional integer sub-lattice called the `fat hook'~\cite{GK11,Kazakov16}.
In the simpler non-graded case ($M=0$), the anti-symmetric fusion can be applied at most $N-1$ times and the fat hook is coincides 
with a sub-lattice in the form of a semi-infinite strip with boundaries 
$s\geq 0$ and $0\leq a \leq N$. On the other hand, there is no restrictions on the anti-symmetric fusion in the graded case provided
$s\leq M-1$. These boundary conditions define an L-shaped sub-lattice within the $(a,s)$-lattice.
\\

There is no unique assignment of the particles to the nodes of the fat hook lattice. The prescription can be made unique
by selecting a particular highest-weight Bethe vacuum which requires an appropriate embedding of the chosen Kac--Dynkin diagram
inside the fat hook. For a general and comprehensive discussion on this we refer the reader to \cite{Volin12}. Below we instead only 
focus on particular physically relevant examples.

\subsection*{Equilibrium ensembles}

A general local equilibrium state in a thermodynamically large system is uniquely characterized by the full set of
functions $\rho_{A}(u)$ pertaining to the densities of Bethe roots of various types, namely the elementary and bound state solutions 
to the (nested) Bethe equations \eqref{eqn:Bethe_equations_scattering}.
Given a complete set of densities, one uniquely specifies a macrostate. The latter correspond to
microcanical ensembles of microstates which are attributed a finite entropy density per mode, reading
\begin{equation}
s_{A}(u) = \rho_{A}(u)\log\left(1+\frac{\ol{\rho}_{A}(u)}{\rho_{A}(u)}\right) +
\ol{\rho}_{A}(u)\log\left(1+\frac{\rho_{A}(u)}{\ol{\rho}_{A}(u)}\right).
\label{eqn:entropy_density}
\end{equation}
We have introduced a set of functions $\ol{\rho}_{A}$, called the hole densities, corresponding to the densities of
the unoccupied solutions to the Bethe equations. Notice that the latter are uniquely determined once $\rho_{A}$ a given,
satisfying the following system of linear integral equations
\begin{equation}
\rho_{A} + \ol{\rho}_{A} = \sigma_{A}K_{A} - K_{AB}\star \rho_{B},
\label{eqn:Bethe-Yang}
\end{equation}
where $\star$ denotes the convolution-type integration over the rapidity domain, define as
\begin{equation}
(f\star g)(u) = \sum_{A}\int \dd w f_{A}(u,w)g_{A}(w),\qquad
(F\star g)(u) = \sum_{B}\int \dd w F_{AB}(u,w)g_{B}(w),
\label{eqn:convolutions_def}
\end{equation}
where the convention over repeated indices has been adopted.
The kernels entering in Eqs.~\eqref{eqn:Bethe-Yang} are the logarithmic derivatives of scattering amplitudes,
\begin{equation}
K_{A}(u) = \frac{1}{2\pi \ii}\partial_{u}\log S_{A}(u),\qquad
K_{AB}(u) = \frac{1}{2\pi \ii}\partial_{u}\log S_{AB}(u).
\label{eqn:kernels_def}
\end{equation}
Finally, the $\sigma$-parity is a $\mathbb{Z}_{2}$ label defined as the sign of the bare momentum derivative,
$\sigma_{A}={\rm sign}(k^{\prime}_{A})$.
Moreover, the total density of the available states for a particle of type $A$ is (due to interactions) a rapidity (i.e. momentum) 
dependent quantity,
\begin{equation}
\rho^{\rm tot}_{A}(u) = \rho_{A}(u) + \ol{\rho}_{A}(u) = \sigma_{A}\frac{p^{\prime}_{A}(u)}{2\pi}.
\label{eqn:total_density_momentum}
\end{equation}

An alternative way to characterize equilibrium state is via the mode occupation functions
\begin{equation}
\vartheta_{A}(u) = \frac{\rho_{A}(u)}{\rho^{\rm tot}_{A}(u)},
\label{eqn:occupation_functions_def}
\end{equation}
which are a natural generalization of the Fermi--Dirac occupation functions to interacting models which are subjected to
the Fermi `exclusion principle'. In the dressing formalism of the Thermodynamic Bethe Ansatz,
another useful set of quantities to define are hole to particle ratios,
\begin{equation}
Y_{A}(u) = \frac{\ol{\rho}_{A}(u)}{\rho_{A}(u)},
\label{eqn:Y_functions_def}
\end{equation}
which are commonly known as the $Y$-functions.
\\

The thermodynamic free energy $f=-\log \mathcal{Z}$ of a general equilibrium state can be conveniently expressed a
functional integral over rapidity distributions,
\begin{equation}
\mathcal{Z} = \int \mathrm{D}[\{\rho_{A}(u)\}]\exp{\left(-L\sum_{A}\int \dd u \big(\mu_{A}(u)\rho_{A}(u) - s_{A}(u)\big)\right)},
\label{eqn:GGE_partition_sum}
\end{equation}
where $s_{A}(u)$ denotes the entropy density per particle given by Eq.~\eqref{eqn:entropy_density}, and $\mu_{A}(u)$ is
a complete set of analytic (dynamical) chemical potentials as defined in~\cite{IQC17}. It is important to stress that
$\mu_{A}(u)$ uniquely define a `generalized' equilibrium ensemble in the sense that the saddle-point of
Eq.~\eqref{eqn:GGE_partition_sum} yields a particular set of Bethe root densities $\rho_{A}(u)$. Alternatively,
a local equilibrium state is fully specified by the expectation values of the local conservation laws~\cite{String_charge}.

In the $L \to \infty$ limit, the saddle-point integration yields a set of coupled non-linear integral equations,
which we refer here to as the canonical TBA equations. In terms of the $Y$-functions defined in Eq.~\eqref{eqn:Y_functions_def},
these are of the form
\begin{equation}
\log Y_{A} = \mu_{A} + K_{AB}\star \log(1+1/Y_{B}).
\label{eqn:TBA_canonical_form}
\end{equation}
Hence, given $\mu_{A}(u)$ one finds $\vartheta_{A}(u)$ via Eq.~\eqref{eqn:TBA_canonical_form}, and vice-versa.
For instance, in the case of the grand-canonical Gibbs ensemble, the chemical potentials read
\begin{equation}
\mu^{\rm GC}_{A}(u) = \beta\,e_{A}(u)+\sum_{i}2h_{i}n_{A,i},
\end{equation}
where $\beta$ is the inverse temperature, $e_{A}(u)$ denoted the one-particle energies,
$\lim_{L\to \infty}\frac{E}{L} = e_{A}\star \rho_{A}$, and $n_{A,i}$ one-particle bare $U(1)$ charges.

\subsection*{Universal dressing transformation}

An infinite sum over particle species on the right-hand side of Eq.~\eqref{eqn:TBA_canonical_form} can be removed by
exploiting certain fusion identities. It is useful to define the \emph{left} inverse of the convolution kernel $(K+1)$,
\begin{equation}
C_{AB}(u) = (K(u)+\delta)^{-1}_{AB}=\delta_{AB}-\fs(u)I_{AB},
\label{eqn:Baxter_Cartan_definition}
\end{equation}
where
\begin{equation}
\fs(u)=\frac{1}{2\cosh{(\pi u)}},
\label{eqn:s-kernel}
\end{equation}
is the solution to equation $K_{1}-\fs\star K_{2}=\fs$, whereas the incidence matrix $I_{AB}$
for the graded $SU(N|M)$-symmetric quantum chains which appears in the Baxter--Cartan matrix
$C_{AB}$ (cf. Eq.~\eqref{eqn:Baxter_Cartan_definition}) splits into the horizontal and vertical parts,
\begin{equation}
I_{s,s^{\prime}} = \delta_{s+1,s^{\prime}}+\delta_{s-1,s^{\prime}},\qquad
I_{a,a^{\prime}} = \delta_{a-1,a^{\prime}}+\delta_{a+1,a^{\prime}},
\end{equation}
respectively, leaving the boundary conditions (which depend on $N$ and $M$) implicit for the moment.

Notice, moreover, that the bare energies of the \emph{fundamental} particles are simply proportional to the elementary
scattering kernels $e_{A}(u) \simeq K_{A}(u)$, whence
\begin{equation}
(K+\delta)^{-1}_{AB}\star K_{B}\equiv C_{AB}\star K_{B}=\delta_{A,1}\fs.
\end{equation}

Taking advantage of the fact that the entire TBA framework originates from the fusion rules for the
Yangian extensions of the classical characters associated to the rectangular Young tableaux,
the entire TBA dressing formalism can in fact be presented in a \emph{universal group-theoretic form}
\begin{equation}
C_{s,s^{\prime}} \star L_{a,s^{\prime}}-C_{a,a^{\prime}}\star \ol{L}_{a^{\prime},s} = \nu_{a,s},
\label{eqn:universal_dressing_abstract}
\end{equation}
where indices $(a,s)\subset \mathbb{Z}^{2}$ belong to the interior of the fat hook lattice.
The actual physical meaning of these equations depends on the interpretation of variables $L_{a,s}$ and $\ol{L}_{a,s}$:
\begin{itemize}
\item for $L_{a,s}\equiv \ol{\rho}_{a,s}$ and $\ol{L}_{a,s}=-\rho_{a,s}$ one finds the Bethe--Yang integral equations
\eqref{eqn:Bethe-Yang}, expressing hole distribution functions $\ol{\rho}_{a,s}$ in terms of Bethe root densities $\rho_{a,s}$, 
with source terms $\nu_{a,s}=k^{\prime}_{a,s}$,
\item for $L_{a,s}\equiv \log(1+Y_{a,s})$ and $\ol{L}_{a,s}\equiv \log(1+Y^{-1}_{a,s})$, one finds the
quasi-local form of TBA equations. Dependence on chemical potentials $\mu_{A}(u)$ is contained in the source terms
\begin{equation}
\nu_{A}=C_{AB}\star \mu_{B}.
\end{equation}
\end{itemize}

In practice, it is useful to use the differential form of Eqs.~\eqref{eqn:universal_dressing_abstract}.
This yields the following system of coupled \emph{linear} integral equations,
\begin{equation}
C^{(\vartheta)}_{AB}\star (q^{\prime}_{B})^{\rm dr} = q^{\prime}_{A},
\end{equation}
or explicitly,
\begin{equation}
C^{(\vartheta)}_{s,s^{\prime}} \star (q^{\prime}_{a,s^{\prime}})^{\rm dr} +
C^{(\vartheta)}_{a,a^{\prime}} \star (q^{\prime}_{a^{\prime},s})^{\rm dr} = q^{\prime}_{a,s},
\end{equation}
where we have introduced the \emph{dressed} Baxter--Cartan matrices
\begin{align}
C^{(\vartheta)}_{s,s^{\prime}}(u) &= \delta_{s,s^{\prime}} - \fs(u)\,I_{s,s^{\prime}}\ol{\vartheta}_{a,s}(u),\\
C^{(\vartheta)}_{a,a^{\prime}}(u) &= \delta_{a,a^{\prime}} - \fs(u)\,I_{a,a^{\prime}}\vartheta_{a,s}(u),
\end{align}
where $\vartheta_{a,s}$ enter as input variables parametrizing the reference many-body vacuum (equilibrium state).
\\

In the non-graded chains ($M=0$), the fat-hook lattice is a semi-infinite strip and the indices range in
$s=1,2,\ldots$ and $a=1\sim N$. In the graded cases (i.e. for $M>0$), the exterior (interior) boundaries are
along $(0,s\geq 0)$ and $(a\geq 0,0)$ ($(N,s\geq M)$ and $(N\geq a,M)$).
There is an additional \emph{exceptional} relation associated to the
boundary node $(a,s)=(N,M)$, which is a particularity of the graded models and
where Eqs.~\eqref{eqn:universal_dressing_abstract} to no apply. Nonetheless, there is no real ambiguity
since the functional relations for quantum characters enforce its uniqueness.

\subsection*{High-temperature expansion}
The TBA dressing equations do not permit closed-form solutions in general. There are two important exceptions to this, however:
(i) the ground-state limit $\beta^{-1}\to 0$ limit and (ii) the high-temperature $\beta \to 0$ limit.
Below we specialize our treatment to the high-temperature limit of the grand canonical Gibbs ensembles where
the dressing integral equations becomes a set of coupled \emph{algebraic} equations, see e.g. \cite{Takahashi1999}.
Here we make use of the group-theoretical formulation by invoking the character formulae for classical (graded) Lie algebras.\\

We begin by the leading-order $\beta$-expansions of the TBA functions,
\begin{align}
\log Y_{A} &= \log Y^{(0)}_{A} + \beta\,F_{A} + \mathcal{O}(\beta^{2}),\\
\log (1+Y_{A}) &= \log (1+Y^{(0)}_{A}) + \beta\,\ol{\vartheta}^{(0)}_{A}F_{A} + \mathcal{O}(\beta^{2}),\\
\log (1+1/Y_{A}) &= \log (1+1/Y^{(0)}_{A}) - \beta\,\vartheta^{(0)}_{A}F_{A} + \mathcal{O}(\beta^{2}).
\label{eqn:temperature_expansion}
\end{align}
The major simplification in the $\beta \to 0$ limit is that the occupation functions become constant (i.e. rapidity-independent) 
functions, and thus all convolution integrals reduce to scalar multiplication.

\subsubsection*{Mode occupation functions}
Using the property of the $s$-kernel, $1\star s = \tfrac{1}{2}$, equations \eqref{eqn:temperature_expansion}
the $\beta \to 0$ limit reduce to the following non-linear functional relations for the $Y$-functions
\begin{equation}
\log \left[Y^{(0)}_{a,s}\right]^{2} = I_{s,s^{\prime}}\log(1+Y^{(0)}_{a,s^{\prime}})-I_{a,a^{\prime}}\log(1+1/Y^{(0)}_{a^{\prime},s}).
\end{equation}
There is an equivalent exponential form of these equations,
\begin{equation}
\left[Y^{(0)}_{a,s}\right]^{2} = \frac{(1+Y^{(0)}_{a,s-1})(1+Y^{(0)}_{a,s+1})}{(1+1/Y^{(0)}_{s,a-1})(1+1/Y^{(0)}_{s,a+1})},
\label{eqn:constant_Y_system}
\end{equation}
for constant (i.e. rapidity-independent) $Y$-functions, which is nothing but the simplified `classical' version of the $Y$-system 
relations~\cite{Zamolodchikov91}.
To find a unique solution to this system we have to solve a system of coupled recurrence relations.
To achieve this, we have to additionally supply the following $N+M-2$ asymptotic conditions,
\begin{align}
\lim_{s\to \infty}Y^{(0)}_{a,s} = \exp{(2h_{a}s)},\qquad a=1 \sim N-1,\\
\lim_{a\to \infty}Y^{(0)}_{a,s} = \exp{(2\mu_{s}a)},\qquad s=1 \sim M-1.
\end{align}
Parameters $h_{a}$ and $\mu_{s}$ are the chemical potentials associated with global conserved
$U(1)$ charges of an equilibrium equilibrium state. In the graded chains, i.e. for $M>1$, there is an additional chemical potential
which is not encoded in the asymptotics of the TBA $Y$-functions. As we shall demonstrate on explicit examples,
the latter enters through the equation associated with the corner node $(a,s)=(N,M)$ of the fat hook.

Any solution to equations \eqref{eqn:constant_Y_system} admits an equivalent gauge-covariant parametrization in terms of classical
characters $\chi_{a,s}$, related to the $Y$-function by a non-linear transformation
\begin{equation}
Y^{(0)}_{a,s}= \frac{\chi_{a,s-1}\chi_{a,s+1}}{\chi_{a-1,s}\chi_{a+1,s}}.
\label{eqn:Y_to_chi}
\end{equation}
The infinite set of functions $\chi_{a,s}$ satisfy the simplified version of the Hirota bilinear relations
\begin{equation}
\chi^{2}_{a,s} = \chi_{a-1,s}\chi_{a+1,s} + \chi_{a,s-1}\chi_{a,s+1}.
\label{eqn:Hirota_classical}
\end{equation}
Indeed, the above formula is just a reduction of the full `quantum' Hirota equation with spectral parameter (see e.g. \cite{Kazakov16}) in the `classical limit' when dependence on the spectral parameter drops out.
In fact, Eq.~\eqref{eqn:Hirota_classical} is the well-known identity for characters $\chi_{a,s}=\chi_{a,s}(G)$ of rectangular 
irreducible representations $[a,s]$ of classical graded algebras $\mathfrak{gl}(N|M)$, with $G$ denoting an element of a
$(N+M-1)$-dimensional Cartan subalgebra. Physically speaking, characters $\chi_{a,s}$ are thus only functions of
the $U(1)$ chemical potentials which parametrize the infinite-temperature grand canonical Gibbs ensemble.
\\

Below we recall some basic facts about the character formulae for the non-graded $\mathfrak{gl}(N)$ Lie algebras,
where only the \emph{rectangular} characters $\chi_{a,s}(G)$, with $G={\rm diag}(x_{1},\ldots,x_{N})$ denoting
a general element of the Cartan subalgebra, will be of our interest.
A character $\chi_{a,s}$ is expressible as a determinant including only the totally symmetric (anti-symmetric)
characters $\chi_{1,s}$ ($\chi_{a,1}$) in accordance with the Giambelli-Jacobi--Trudi formula
\begin{equation}
\chi_{a,s}(G) = {\rm Det}\,\left(\chi_{1,s+j-k}\right)_{1\leq j,k\leq a}.
\end{equation}
An explicit parametrization in terms of the eigenvalues of the Cartan charges is given by the $1$st Weyl character formula
\begin{equation}
\chi_{a,s}(G) = \frac{{\rm Det}\,(x^{N-j+s\,\theta_{a,j}}_{k})_{1\leq j,k\leq N}}{{\rm Det}\,(x^{N-j}_{k})_{1\leq j,k\leq N}},
\end{equation}
with $\theta_{i,j}=1$ if $i\geq j$ and zero otherwise. Functions $\chi_{a,s}(G)$ are related to Schur polynomials, i.e. completely
symmetric polynomials of $N$ variables $x_{1},\ldots,x_{N}$. The generating function for totally symmetric characters is
\begin{equation}
w(z) = \prod_{j=1}^{N}\frac{1}{1-z\,x_{j}} = \sum_{s=0}^{\infty}z^{s}\chi_{1,s}(x_{1},\ldots,x_{N}).
\end{equation}
Likewise, the totally anti-symmetric characters are generated from the inverse expansion
$w^{-1}(z) = \sum_{a=1}^{\infty}(-1)^{a}\chi_{a,1}z^{a}$.
For instance, in the simplest $N=2$ case we have
\begin{equation}
\chi_{1,s}(x_{1},x_{2}) =
\frac{
\begin{vmatrix}
x^{s+1}_{1} & x^{s+1}_{2} \\
1 & 1
\end{vmatrix}
}
{
\begin{vmatrix}
x_{1} & x_{2} \\
1 & 1
\end{vmatrix}
} =
\frac{x^{s+1}_{1} - x^{s+1}_{2}}{x_{1}-x_{2}}.
\end{equation}
The Cartan sector of the $SU(2)$-symmetric fundamental spin chain is thus parametrized by a single parameter $h$,
determined by $x_{1}/x_{2}=\exp{(2h)}$, which pertains to the chemical potential for the conserved total spin $\hat{S}^{z}$.
Therefore $\chi_{0,s}=\chi_{2,s}=1$, implying
\begin{align}
1+Y^{(0)}_{s}(h) = \left[\chi_{1,s}(h)\right]^{2},
\end{align}
with symmetric characters
\begin{equation}
\chi_{1,s}(h) = \frac{e^{-(s+1)h}-e^{(s+1)h}}{e^{-h}-e^{h}}.
\end{equation}
In the limit of half filling, i.e. $h\to 0$, we have in particular
\begin{align}
\lim_{h\to 0}\chi_{1,s}(h) &= d_{s} = {\rm dim}\,V_{s}=s+1,\\
\lim_{h\to 0}Y^{(0)}_{s}(h) &= s(s+2).
\end{align}

The higher-rank $\mathfrak{su}(N)$-symmetric models involve $N-1$ conserved number operators $\hat{N}_{i}$.
We thus put
\begin{equation}
x_{1}=1,\qquad x_{j}/x_{j+1}=\exp{(2h_{j})}.
\end{equation}

Finally, it is instructive to examine the singular limit of vanishing chemical potentials $x_{j}\to 1$.
For a general partition (Young tableaux) $\lambda=(\lambda_{1},\ldots,\lambda_{N})$, with
$\lambda_{j}\geq \lambda_{j+1}$ defining a $\mathfrak{gl}(N)$ representation, the latter corresponds to
the following specialization of Schur polynomials
\begin{equation}
s_{\lambda}(1,1,\ldots,1)=\prod_{1\leq i<j\leq N}\frac{\lambda_{i}-\lambda_{j}+j-i}{j-i},
\label{eqn:Schur_specializatoin}
\end{equation}
which is the well-known hook-length formula yielding the multiplet dimension ${\rm dim}V_{\lambda}$.
For the rectangular partitions $\lambda=(s^{a})=(s,s,\ldots,s)$ this implies $\lim_{h_{j}\to 0}\chi_{a,s}={\rm dim}\,V_{a,s}$.\\

\subsubsection*{Dressing in the high-temperature limit}

In the infinite temperature limit $\beta \to 0$, the dressing transformation becomes an infinite system of coupled algebraic
equations which admits a closed-form solution.
The leading high-temperature contribution to the mode occupation functions and
the dressed values of the $U(1)$ charges can be directly computed from the TBA $Y$-functions, as outlined below.
Specifically, the occupation functions are expressible as the following ratios of $\chi$-functions
\begin{equation}
\vartheta^{(0)}_{a,s} = \frac{\chi_{a-1,s}\chi_{a+1,s}}{\chi^{2}_{a,s}},\qquad
\ol{\vartheta}^{(0)}_{a,s} = \frac{\chi_{a,s-1}\chi_{a,s+1}}{\chi^{2}_{a,s}}.
\label{eqn:occupation_functions_characters}
\end{equation}
The dressed values of conserved $U(1)$ charges are most easily computed from
\begin{align}
m^{\rm dr (0)}_{a,s}(G) &= \partial_{2h_{a}}\log Y^{(0)}_{a,s}(G),\qquad a=1\sim N-1,\\
n^{\rm dr (0)}_{a,s}(G) &= \partial_{2\mu_{s}}\log Y^{(0)}_{a,s}(G),\qquad s=1\sim M-1,
\end{align}
where $G$ depends on a set of chemical potentials $h_{i}$, $\mu_{i}$. There is an extra chemical potential, denoted by $\uu$,
which is not encoded in the asymptotic but explicitly enters in the equation for the corner node.

The high-temperature limit of the particles' dressed dispersion relations in the leading order
$\mathcal{O}(\beta)$ is found as a solution to the following system of coupled linear integral equations
\begin{equation}
F_{a,s} = C^{(0)}_{s,s^{\prime}}\star F_{a,s^{\prime}} + C^{(0)}_{a,a^{\prime}}\star F_{a^{\prime},s} - \nu_{a,s},
\label{eqn:F_system}
\end{equation}
where $C^{(0)}_{a,a^{\prime}}$ and $C^{(0)}_{s,s^{\prime}}$ are the Baxter--Cartan matrices
dressed by the \emph{infinite-temperature} equilibrium state,
\begin{align}
C^{(0)}_{s,s^{\prime}}\star F_{a,s^{\prime}} &= \delta_{s,s^{\prime}} -
I_{s,s^{\prime}} \fs \star \ol{\vartheta}^{(0)}_{a,s^{\prime}}F_{a,s^{\prime}},\\
C^{(0)}_{a,a^{\prime}}\star F_{a^{\prime},s} &= \delta_{a,a^{\prime}} -
I_{a,a^{\prime}}\fs \star \vartheta^{(0)}_{a^{\prime},s} F_{a^{\prime},s}.
\end{align}

\subsection*{Anisotropic Heisenberg spin-$\frac{1}{2}$ chain}

A prototype model of an integrable spin chain is the axially anisotropic Heisenberg spin-$1/2$ chain (XXZ model),
\begin{equation}
\hat{H}_{\rm XXZ} \simeq \sum_{j=1}^{L}\left(\hat{S}^{x}_{j}\hat{S}^{x}_{j+1}+\hat{S}^{y}_{j}\hat{S}^{y}_{j+1}+
\Delta \hat{S}^{z}_{j}\hat{S}^{z}_{j+1}\right).
\label{eqn:XXZ_Hamiltonian}
\end{equation}
For $\Delta = 1$, there is manifest global $SU(2)$ non-Abelian symmetry. However, integrability implies a hidden
(i.e. non-manifest) infinite dimensional quantum-group symmetry algebra known as the Yangian $\mathcal{Y}(\mathfrak{su}(2))$.
For generic values $\Delta \neq 1$ the symmetry continuously deforms in the so-called `quantum deformed' universal enveloping algebra
$\mathcal{U}_{q}(\mathfrak{su}(2))$. For the root-of-unity value $q=\pi(m/\ell)$, with co-prime integers $m<\ell$ and $\ell\geq 2$, 
the symmetry enlarges (see \cite{Takahashi1999}).

\subsubsection*{Isotropic point}
We first consider the isotropic point $|\Delta|=1$ regime, the particle content consists of magnons ($s=1$) and an infinite
sequence of magnonic bound states ($s\geq 2$). These particles are commonly referred to as the $s$-strings.
The high-temperature limit of the dressing transformation is a three-point recurrence relation
\begin{equation}
C^{(0)}_{s,s^{\prime}}\star F_{s^{\prime}} = \delta_{s,1}\fs,\qquad
\lim_{s\to \infty}F_{s}=0.
\label{eqn:su2_recurrence}
\end{equation}

The occupation functions are readily obtained from the character formulae as prescribed by 
Eqs.~\eqref{eqn:occupation_functions_characters},
\begin{align}
\vartheta^{(0)}_{s}(h) &= \frac{1}{1+Y_{s}(h)} = \frac{1}{\chi^{2}_{s}(h)},\\
\ol{\vartheta}^{(0)}_{s}(h) &= (1+1/Y_{s}(h))^{-1} = 1-\vartheta_{s}(h),\\
Y_s (h) &= \frac{\sinh((s+1)h)^2}{\sinh(h)^2} -1. 
\end{align}
The solution to the recurrence relation \eqref{eqn:su2_recurrence} reads
\begin{equation}
F_{s}(h) = \frac{\chi_{s}(h)}{\chi_{1}(h)}\left(\frac{K_{s}}{\chi_{s-1}(h)}-\frac{K_{s+1}}{\chi_{s+1}(h)}\right).
\end{equation}
In particular, exactly at half filling we have
\begin{equation}
\lim_{h\to 0} \ol{\vartheta}^{(0)}_{s}(h) = \frac{s(s+2)}{(s+1)^{2}},\qquad
\lim_{h\to 0}F_{s}(h) = \frac{s+1}{2}\left(\frac{K_{s}}{s}-\frac{K_{s+2}}{s+2}\right).
\end{equation}
The dressed momentum in the $\beta \to 0$ limit therefore reads
\begin{equation}
p^{\prime (0)}_{s}(u) = \frac{s+1}{2}\left(\frac{K_{s}(u)}{s} - \frac{K_{s+2}(u)}{s+2}\right).
\end{equation}
Similarly, the dressed energy is of the form
\begin{equation}
\varepsilon^{\prime (0)}_{s}(u) = \lim_{\beta \to 0}\beta^{-1}\partial_{u}\log Y_{s}(u) = F^{\prime}_{s}(h)
= \frac{s+1}{2}\left(\frac{K^{\prime}_{s}(u)}{s} - \frac{K^{\prime}_{s+2}(u)}{s+2}\right).
\end{equation}

The dressed spin is obtained from the solution of the following homogeneous recurrence
\begin{equation}
m^{\rm dr(0)}_{s}(h) - \frac{1}{2}I_{s,s^{\prime}}\ol{\vartheta}^{(0)}_{s^{\prime}}(h)m^{\rm dr(0)}_{s^{\prime}}(h) = 0,\qquad
\lim_{s\to \infty}m^{\rm dr(0)}_{s}(h)=s.
\end{equation}
For finite value of chemical potential $h$, the solution reads
\begin{equation}
m^{\rm dr(0)}_{s}(h) = \partial_{2h}\log Y^{(0)}_{s}(h) =
\frac{\sinh{(h)}}{\sinh{((s+1)h)}}\left(\frac{s}{\sinh{(s\,h)}}-\frac{s+2}{\sinh{((s+2)h)}}\right).
\end{equation}
In the vicinity of half filling $h\to 0$ we thus have
\begin{equation}
\qquad m^{\rm dr(0)}_{s}(h) \sim \frac{1}{3}(s+1)^{2}h + \mathcal{O}(h^{3}).
\end{equation}

\subsubsection*{Easy-axis (gapped) regime $|\Delta|>1$}
The so-called `easy-axis' regime of the XXZ Hamiltonian \eqref{eqn:XXZ_Hamiltonian} is parametrized by anisotropy
$\Delta = \cosh{(\eta)}$, for $\eta \in \mathbb{R}_{+}$. The elementary kernels undergo a trigonometric deformation
\begin{equation}
K_{s}(u) = \frac{1}{2\pi \ii}\partial_{u}\log S(u) = \frac{1}{2\pi \ii}\frac{2\sinh{(s\eta)}}{\cosh{(s\eta)}-\cos{(2u)}}.
\end{equation}
The fundamental zone of the rapidity integration is a compact interval $u\in [-\eta/2,\eta/2]$. The kernels for the
isotropic chain are recovered in the limit $\eta \to 0^{+}$ after simultaneous rescaling the rapidity variable,
\begin{equation}
K^{\rm XXX}_{s}(u) = \lim_{\eta \to 0^{+}}K_{s}(\eta u)= \frac{1}{2\pi}\frac{s}{(s/2)^{2}+u^{2}}.
\end{equation}
The bare energies of the $s$-strings are
\begin{equation}
e_{s}(u) = -\pi \sinh{(\eta)}K_{s}(u).
\end{equation}
In the large-$s$ limit we the kernels behave as
\begin{equation}
\lim_{s\to \infty}K_{s}(u) = \frac{1}{\pi},\qquad
\lim_{s\to \infty}K^{\prime}_{s}(u) = \frac{\sin{(2u)}}{\pi \cosh{(s\eta)}}.
\end{equation}
The rapidity derivatives of the dressed energies in the high-temperature limit at half filling read
\begin{equation}
\varepsilon^{\prime (0)}_{s}(u) = -\pi \sinh{(\eta)} \frac{s+1}{2s(s+2)}\left((s+2)K^{\prime}_{s}-sK^{\prime}_{s+2}\right).
\end{equation}
The expressions for the rapidity-independent quantities $m^{\rm dr (0)}_{s}(h)$ and $\vartheta^{(0)}_{s}(h)$ are the same as those
for the isotropic point.

\subsubsection*{Easy-plane (gapless) regime $|\Delta|<1$}
The easy-plane regime is parametrized by $\Delta=\cos{(\gamma)}$, for $\gamma \in \mathbb{R}$.
We consider the root-of-unity value of the quantum deformation parameter $q=e^{\ii \gamma}$, namely $\gamma$ which are
rational multiples of $\pi$, $\gamma/\pi = m/\ell$, in which case the centre of the quantum group symmetry algebra
$\mathcal{U}_{q}(\mathfrak{sl}_{2})$ enlarges.
The structure of eigenstates and thermodynamic particle content becomes dependent on $\Delta$ in a rather intricate
way. The complete classification can be found in \cite{Takahashi1999}. A key difference in compare to the
$|\Delta|\geq 1$ regime is that the number and types of excitations in the spectrum now explicitly depends on the value of $\gamma$.
To further simplify the analysis, we restrict our consideration to a discrete set of primitive roots of unity $\gamma = \pi/\ell$,
for $\ell \in \mathbb{N}$ ($\ell \geq 2$), when there are $\ell$ distinct species: for $s=1\sim \ell-1$ we have
the magnons and bound states thereof with bare spin $n_{s}=s$, whereas the `last particle' labelled by $s=\ell$ corresponds to
an unbound magnon ($n_{\ell}=1$) of negative $\sigma$-parity ($\sigma_{\ell}=-1$).
The origin of such a truncation can be traced to the fact that at these particular values of $q$ the $(\ell+1)$-dimensional
irreducible representation of $\mathcal{U}_{q}(\mathfrak{sl}(2))$ becomes reducible.
The last two particles interact with other particles in a distinctive way. This special feature of the 
gapless regime is key to understand the nature of quantum spin transport \cite{IN_Drude}.

The TBA $Y$-functions will be denoted by $Y_{s}=\ol{\rho}_{s}/\rho_{s}$ for $s=1\sim \ell-2$, whereas the special pair
of particles is associated the following $Y$-functions
\begin{equation}
Y_{\ell-1}\equiv Y_{\circ}=\ol{\rho}_{\circ}/\rho_{\circ},\qquad
Y_{\ell}\equiv Y_{\bullet}=\rho_{\bullet}/\ol{\rho}_{\bullet}.
\end{equation}
The elementary scattering kernels depend on $\gamma$ are read explicitly
\begin{equation}
K_{s}(u) = \frac{2\sin{(\gamma q_{j})}}{\cosh{(2u)}+\cos{(\gamma q_{j})}},
\end{equation}
with $q_{j}=\ell-j$ for $j=1\sim \ell-1$ and $q_{\ell}=-1$, and satisfying the following identities
\begin{align}
K_{s} - \fs\star (K_{s-1}+K_{s+1}) &= 0,\qquad s\leq \ell-2,\\
K_{\circ} = -K_{\bullet} &= \fs \star K_{\ell-2},
\end{align}
The $\fs$-kernel gets modified and now depends on the anisotropy parameter $\ell$,
\begin{equation}
\fs_{\ell}(u)=\frac{\ell}{2\pi}\frac{1}{\cosh{(2u)}}.
\end{equation}

The quasi-local form of the Bethe--Yang equations for the string densities is compactly written as
\begin{equation}
\rho^{\rm tot}_{s} = I^{(\ell)}_{s,s^{\prime}}\star \ol{\rho}_{s^{\prime}},
\end{equation}
where $I^{(\ell)}$ stands for the $\ell$-dimensional incidence matrix of the $D_{\ell}$ system,
\begin{equation}
I^{(\ell)}_{jk}=\sum_{j=1}^{\ell-3}(\delta_{j,k-1}+\delta_{j,k+1})
+ \delta_{\ell-2,\ell}  + \delta_{\ell-1,\ell-2} + \delta_{\ell,\ell-2}.
\end{equation}
Similarly, the quasi-local TBA equation take the form
\begin{equation}
\log Y_{s} = \nu_{s} + I_{s,s^{\prime}}\fs \star \log(1+Y_{s^{\prime}}).
\end{equation}
Notice that $\nu_{\circ,\bullet}=\nu_{\circ,\bullet}(\ell,h)$.

Unlike in the isotropic chain or the gapped phase, the dressing equations in gapless regime for the root-of-unity values of
$\gamma$ enclose a \emph{finite} set of equations
\begin{align}
(1+1/Y_{s})f_{s} - \fs_{\ell} \star (f_{s-1}+f_{s+1}) &= -\delta_{s,1}\fs_{\ell},\qquad s=1\sim \ell-3,\\
(1+1/Y_{\ell-2})f_{\ell-2} - \fs_{\ell} \star(f_{\ell-3} + 2f_{\circ}) &= 0,\\
(1+1/Y_{\circ})f_{\circ} - \fs_{\ell} \star f_{\ell-2} &= 0,
\label{eqn:gapless_Taka}
\end{align}
where we denoted $f_{s}\equiv \ol{\vartheta}_{s}F_{s}$.
The general solution of equations \eqref{eqn:gapless_Taka} is already known~\cite{Takahashi1999}
\begin{align}
\hat{f}_{s}(\kappa) &= \frac{1}{\chi_{1}\chi_{y_{1}-1}\chi_{s+y_{1}-1}}
\left(\chi_{s+2y_{1}-1}\frac{\sinh{(\tfrac{\pi}{2}\tfrac{q_{s}}{p_{0}}\kappa)}}{\sinh{(\tfrac{\pi}{2}\kappa)}}-
\chi_{s-1}\frac{\sinh{(\tfrac{\pi}{2}\tfrac{q_{s}-p_{1}}{p_{0}}\kappa)}}{\sinh{(\tfrac{\pi}{2}\kappa)}}\right),\quad s=1\sim \ell-2,\\
\hat{f}_{\circ}(\kappa) &= \frac{\sinh{(\tfrac{\gamma}{2}\kappa)}}{2\sinh{(\tfrac{\pi}{2}\kappa)}},
\end{align}
where
\begin{align}
y_{1} &= 1,\quad p_{0} = \ell,\quad p_{1} = 1,\\
q_{s} &= \ell-s,\quad s=1\sim \ell-1,\qquad q_{\ell} = -1,
\end{align}
are the so-called Takahashi--Suzuki numbers for the simple roots of unity $q=\cos{(\pi/\ell)}$.
In Fourier space the kernel read
\begin{equation}
\hat{K}_{s}(\kappa) = \frac{\sinh{(\tfrac{\gamma}{2}q_{s}\kappa)}}{\sinh{(\tfrac{\pi}{2}\kappa)}},\qquad
\hat{\fs}(\kappa) = \frac{1}{2\cosh{(\tfrac{\gamma}{2}\kappa)}}.
\end{equation}

The regular particles ($s$-strings) with indices ranging in $s=1\sim \ell-2$ behave similarly to
the regular $s$-strings in the isotropic and gapped regimes $|\Delta|\geq 1$, and obey the standard $Y$-system relations
\begin{equation}
\left[Y^{(0)}_{s}(h)\right]^{2} = \left(1+Y^{(0)}_{s-1}(h)\right)\left(1+Y^{(0)}_{s+1}(h)\right),\qquad
s=1\sim \ell-3.
\label{eqn:regular_Y_system_gapless}
\end{equation}
On the other hand, the pair of exceptional excitations which are due to the truncated particle spectrum require a separate
analysis. For their $Y$-functions we find
\begin{equation}
\log \left[\frac{Y_{\circ}(h)}{Y_{\bullet}(h)}\right]=2h\ell,
\end{equation}
which is valid for any value of inverse temperature $\beta$. We thus parametrize
\begin{equation}
Y_{\bullet}(h)=e^{-2h\ell}Y_{\circ}(h),
\end{equation}
which implies that in the $h\to 0$ limit $\lim_{h\to 0}(Y_{\circ}/Y_{\bullet})=1$.
Similarly, their total densities also coincide $\rho^{\rm tot}_{\circ}=\rho^{\rm tot}_{\bullet}$ (for any $\beta$).
whereas the dressed momenta differ by an overall sign, $p^{\prime}_{\circ}=-p^{\prime}_{\bullet}$,
due to opposite $\sigma$-parities $\sigma_{\circ}=-\sigma_{\bullet}$.
In the $\beta \to 0$ limit and finite $h$, the $Y$-system functional
relations \eqref{eqn:regular_Y_system_gapless} for the regular particles together with
\begin{align}
\left[Y^{(0)}_{\ell-2}(h)\right]^{2} &= \left(1+Y^{(0)}_{\ell-3}(h)\right)\left(1+Y^{(0)}_{\circ}(h)\right)
\left(1+Y^{(0)}_{\bullet}(h)\right), \nonumber \\
\left[e^{-h\ell}Y^{(0)}_{\circ}(h)\right]^{2} &= \left[e^{h\ell}Y^{(0)}_{\bullet}(h)\right]^{2} = 1+Y^{(0)}_{\ell-2}(h),
\end{align}
enclosing a finite system of $\ell$ algebraic relations, with the solution
\begin{align}
Y^{(0)}_{s}(h) &= \chi^{2}_{1,s}(h)-1,\qquad s=1\sim \ell-2,\\
Y^{(0)}_{\circ}(h) &= e^{h\,\ell}\chi_{1,\ell-2}(h).
\end{align}
At half filling $h\to 0$, the high-temperature regular $Y$-functions become
\begin{equation}
\lim_{h\to 0}Y^{(0)}_{s}(h) = s(s+2),\qquad s=1\sim \ell-2,
\end{equation}
and the regular particle carry the dressed spin
\begin{equation}
m^{\rm dr(0)}_{s}(h) \sim \frac{1}{3}(s+1)^{2}h + \mathcal{O}(h^{3}),
\end{equation}
which agrees with the results found earlier for the $s$-string excitations in $|\Delta|\geq 1$ regime.
For the special pair of excitations, the mode occupation functions at half filling depend on $\ell$ are read
\begin{equation}
\lim_{h \to 0}Y^{(0)}_{\circ,\bullet}(h) = \ell-1,\qquad
\lim_{h \to 0}\vartheta^{(0)}_{\circ,\bullet}(h) = \frac{1}{\ell}.
\end{equation}
Their dressed energies and dressed spin are computed as
\begin{align}
\varepsilon^{\prime (0)}_{\circ}(u) = \varepsilon^{\prime (0)}_{\bullet}(u) =
\lim_{\beta \to 0}\beta^{-1}\partial_{u}\log Y_{\circ}(u) = \lim_{\beta \to 0}\partial_{u}(\fs \star \log(1+Y_{\ell-2}))
= \partial_{u}(\fs \star f_{\ell-2}) = (1+1/Y^{(0)}_{\circ})f^{\prime}_{\circ},
\end{align}
where we have used the expansion $\log(1+Y_{\ell-2})=\log(1+Y^{(0)}_{\ell-2})+\beta f_{\ell-2}+\mathcal{O}(\beta^{2})$.
In the $h\to 0$ limit, the dressed spin of the special particles reads
\begin{equation}
m^{\rm dr(0)}_{\circ}(h) = -m^{\rm dr(0)}_{\bullet}(h) =
\partial_{2h}\log Y^{(0)}_{\circ}(h) \sim \pm \frac{\ell}{2} + \mathcal{O}(h).
\end{equation}
\\

\paragraph*{Free fermonic point (XX spin chain).}
The non-interacting point corresponds to $\ell = 2$. In this case there is no `regular strings' in the spectrum which now only
comprises the exceptional species $\circ$ and $\bullet$.
Due to the absence of interactions, the scattering kernel simplify to $K_{\circ \circ}=K_{\bullet \bullet}=K_{(2,+)}=0$
and $K_{\circ \bullet}=K_{\bullet \circ}= K_{(2,-)}=0$, and the non-interacing TBA equations read
\begin{equation}
\log Y_{\circ} = 2h -2\pi\beta\,K_{1},\qquad \log Y_{\bullet} = -2h - 2\pi\beta\,K_{1}.
\end{equation}
implying that the dressed spin is just the bare spin,
$\lim_{h\to 0}s^{\rm dr}_{\circ,\bullet}(h)=\partial_{2h}\log Y_{\circ,\bullet}(h)|_{h=0}=1$, as expected.
Indeed, the two particles with $\sigma$-parties $\sigma_{\circ}=1$ and $\sigma_{\bullet}=-1$ at the non-interacting point
$\gamma = \pi/2$ are nothing but two branches of a single free electon dispersion with momentum ranges
$k\in [-\tfrac{\pi}{2},\tfrac{\pi}{2}]$ and $k\in [-\pi,-\tfrac{\pi}{2}]\cup [\tfrac{\pi}{2},\pi]$.

\subsection*{Fermionic chains}

Integrable Hamiltonians given by Eq.~\eqref{eqn:graded_Hamiltonians} with non-trivial grading
are most naturally expressed in terms of canonical fermions.
Below we consider a few most prominent examples which play an important role in the condensed matter literature.

Before proceeding we would like to given a few general technical remarks.
In distinction to the ordinary (non-graded) semi-simple Lie algebras, the $\mathbb{Z}_{2}$-graded Lie algebras
exhibit certain special features. To begin with, there is no unique choice
of simple root system which affects the algebraic Bethe ansatz diagonalization procedure in the sense that the bare (Bethe) vacuum
is no longer unique. This is to be contrasted with the non-graded spin chain where, for instance in the $SU(N)$-symmetric models,
all distinct bare vacua share the same form of Bethe equations modulo particle relabelling.
In the graded models considered here, the total number of different bare vacua available equals
the rank of $\mathfrak{g}=\mathfrak{su}(N|M)$, i.e. ${\rm rank}(\mathfrak{g})=N+M-1$.
Different choices are represented by their corresponding Kac--Dynkin diagrams which consist of $N+M-1$ nodes (each
belonging to an elementary excitation in the spectrum), with the convention that open (bosonic) circles 
correspond to adjacent states of equal Grassmann parity, while crossed (fermionic) circles to adjacent states of opposite parities.
All distinct possibilities are nevertheless interrelated by the so-called fermionic duality transformations (cf. \cite{Volin12})
and permit to construct the same complete spectrum of (highest-weight) eigenstates.
In the fundamental chains there is only one type of elementary excitations which carries momentum and energy,
while the remaining excitations pertain to internal degrees of freedom and are referred to as the auxiliary particles.

Finally, we wish to emphasize that the notion of the momentum-carrying elementary particles and the assignment of the auxiliary 
excitations and bound states thereof depend explicitly on the choice of grading, the high-temperature $Y$-functions $Y^{(0)}_{a,s}$ 
attached to the interior nodes of the fat hook lattice always stay the same.

\subsubsection*{Fermi--Hubbard model}
The one-dimensional Fermi--Hubbard model comprises spin-full electrons which interact via Coulomb repulsion,
\begin{equation}
\hat{H}_{\rm H} = -\sum_{j=1}^{L}\sum_{\sigma \in \ua,\da}\hat{c}^{\dagger}_{j,\sigma}\hat{c}_{j+1,\sigma}
+4\uu\sum_{j=1}^{L}\hat{V}^{\rm H}_{j,j+1},
\label{eqn:Hubbard_Hamiltonian}
\end{equation}
with Hubbard interaction
\begin{align}
\hat{V}^{\rm H}_{j,j+1} = \sum_{j=1}^{L}\left(\hat{n}_{j,\ua}-\tfrac{1}{2})(\hat{n}_{j,\da}-\tfrac{1}{2}\right).
\label{eqn:Hubbard_interaction}
\end{align}
Strictly speaking, the model is not a member of a parameter-less family of $SU(N|M)$-symmetric Hamiltonians
\eqref{eqn:graded_Hamiltonians}. Indeed, the Fermi--Hubbard chain has quite a special place among Bethe Ansatz solvable models as
the underlying quantum algebra which governs the structure of eigenstates is related to a certain the degenerate limit of an 
exceptional central extension of $\mathfrak{su}(2|2)$ graded Lie algebra~\cite{FQ12}.

There are four states per lattice site, $\{\ket{\emptyset},\ket{\ua},\ket{\da},\ket{\bullet}\}$, were
$\ket{\emptyset}$ denotes an empty site and $\ket{\bullet}$ a doubly-occupied site.
There are two types of elementary excitations which constitute two bosonic $\mathfrak{su}(2)$ subalgebras:
$\{\ket{\ua},\ket{\da}\}$ are the spin degrees of freedom,
while $\{\ket{\emptyset},\ket{\bullet}\}$ constitute the charge ($\eta$-spin) $\mathfrak{su}(2)$ degrees of freedom.
The bosonic generators of the spin and charge $\mathfrak{su}(2)$ subalgebras are
\begin{equation}
\hat{S}^{\alpha}=\sum_{j=1}^{L}\hat{S}^{\alpha}_{j},\qquad \hat{\eta}^{\alpha}=\sum_{j=1}^{L}\hat{\eta}^{\alpha}_{j},
\end{equation}
for $\alpha \in \{z,+,-\}$.
The Cartan generators of the global $U(1)$ spin and charge in terms of local electron number operators,
$\hat{n}_{\ua} = \ket{\ua}\bra{\ua}+\ket{\bullet}\bra{\bullet}$ and
$\hat{n}_{\ua} = \ket{\da}\bra{\da}+\ket{\bullet}\bra{\bullet}$,
reading explicitly
\begin{equation}
\hat{S}^{z}_{j} = \tfrac{1}{2}(\hat{n}_{j,\ua}-\hat{n}_{j,\da}),\qquad
\hat{\eta}^{z}_{j} = \tfrac{1}{2}(\hat{n}_{j,\ua}+\hat{n}_{j,\da}-1).
\end{equation}
The local electron number operator $\hat{n}_{\rm e}=\hat{n}_{\ua}+\hat{n}_{\da}$ and the total
number of electrons on an $L$-site lattice is $\hat{N}_{\rm e}=2\hat{\eta}^{z}+L$.

Despite the exceptional status of the Hubbard model, its spectrum may still be embedded in the previously described
universal description of the $SU(N|M)$-symmetric models.
A few modifications are necessary though: the elementary $S$-matrices become a function of the coupling strength parameter $\uu$,
reading
\begin{equation}
S_{n}(u) = \frac{u-n\,\ii \uu}{u+n\,\ii \uu},.
\end{equation}
while the corresponding $s$-kernel acquires $\uu$-dependence
\begin{equation}
\fs(u) = \frac{1}{4\uu \cosh{(\tfrac{\pi}{2}\tfrac{u}{\uu})}}.
\end{equation}
The fused amplitudes for the scattering of $u$ and $w$-roots are
\begin{equation}
S_{n|uw,m|uw}(u)=S^{-1}_{n|w,m|w}(u)=S^{-1}_{nm}(u),
\label{eqn:scattering_amplitudes_Hubbard}
\end{equation}
whereas for the two-branched $y$-particle we have $K_{\pm,n}(u)=\tfrac{1}{2\pi \ii}\partial_{u}\log S_{n}(u)$,
where $n$ is the integer index of either a $s|w$-string or a $a|uw$-stack.

Bethe eigenstates in a finite-volume are characterized in terms of rapidity sets which are solutions to the Lieb--Wu equations
\begin{align}
e^{\ii k(u_{j})L}\prod_{j=1}^{N_{w}}S_{1}(u_{k},w_{j}) &= 1,\\
\prod_{j=1}^{N_{u}}S^{-1}_{1}(w_{k},u_{j})\prod_{j=1}^{N_{w}}S_{1,1}(w_{k},w_{j}) &= -1,
\end{align}
with $2N_{w}\leq N_{u}\leq L$, with the bare electron dispersion reading $u_{j}=\sin{(k_{j})}$. This means that
each Bethe root $u_{j}$ yields two distinct values of momenta $k_{j}$. To this end it is therefore useful
to introduce a double-branched $y$-roots by virtue of the Zhukovsky
transformation $u_{j}=\tfrac{1}{2}(y_{j}+y^{-1}_{j})$; the two branches are given by~\cite{FQ12}
\begin{equation}
y_{\pm}(u)=x(u)^{\pm 1}=x(u+\pm \ii 0),\qquad x(u)=u+u\sqrt{1-1/u^{2}}.
\end{equation}
Note that $x(u)$ has a square-root branch cut along the interval $[-1,1]$.

Although the global symmetry of the Fermi--Hubbard is $SO(4)$, the thermodynamic particle content is related to the
local quantum symmetry and presently complies with fat hook lattice associated to $\mathfrak{g}=\mathfrak{su}(2|2)$.
The assignment of particles to its nodes goes as follows:
\begin{enumerate}
\item The momentum-carrying unbound electrons, pertaining to the two-branched $y$-particles,
$y_{\pm}(u)$ ($u\in [-1,1]$), are attached to the master node at $(1,1)$ and the corner node at $(2,2)$,
\item the auxiliary bound states of spin excitations forming regular $s$-strings are attached to nodes $(1,s+1)$,
\item and the momentum-carrying $a|uw$-stacks, representing spin-singlet bound states composed of $2a$ electrons and $a$
spin-down excitations, are arranged along the nodes $(a+1,1)$.
\end{enumerate}

The canonical TBA equations are of the form
\begin{align}
\log Y_{y} &= \mu_{y} + K_{M}\star \log(1+Y_{M|uw}) - K_{M}\star \log(1+Y_{M|w}),\\
\log Y_{M|uw} &= \mu_{M|uw} + K_{MN}\star \log(1+1/Y_{N|uw}) - K_{M}\hstar \log(1+1/Y_{-}) + K_{M}\hstar \log(1+1/Y_{+}),\\
\log Y_{M|w} &= \mu_{M|w} + K_{MN}\star \log(1+1/Y_{N|w}) - K_{M}\hstar \log(1+1/Y_{-}) + K_{M}\hstar \log(1+1/Y_{+}),
\label{eqn:Hubbard_canonical_TBA}
\end{align}
The canonical source terms depend on the bare energies and $U(1)$ chemical potentials and read
\begin{equation}
\mu_{y}(u) = \beta\,e_{y}(u) - \mu - h,\qquad
\mu_{a|uw}(u) = \beta\,e_{a|uw}(u) - 2a\,\mu,\qquad
\mu_{s|w}(u) = 2s\,h.
\end{equation}
The bare energies of momentum-carrying excitations are
\begin{align}
e_{\pm}(u) &= -2\cos{p_{\pm}(u)} + 2\uu = \pm 2\sqrt{1-u^{2}}+2\uu,\\
e_{a|uw}(u) &= e_{+}(u+M\ii \uu) + e_{-}(u-a\ii \uu) = 2\sqrt{1-(u+a\ii \uu)^{2}} + 2\sqrt{1-(u-a\ii \uu)^{2}},
\end{align}
and $e_{s|w}=0$.
One can get rid off the infinite sums in the last two equations in \eqref{eqn:Hubbard_canonical_TBA} by convolving with
respect to the Baxter--Cartan matrix $C$. Using the property
$C_{aa^{\prime}}\star \mu_{a^{\prime}|uw}=\delta_{a,1}\beta\,\fs \hstar (e_{+}-e_{-})$, one finds the quasi-local TBA equations
\begin{align}
\log Y_{\pm} &= \fs\star \log(1+Y_{1|uw}) - \fs \star \log(1+Y_{1|w}),\\
\log Y_{s|w} &= I_{s,s^{\prime}} \fs \star \log(1+Y_{s^{\prime}|w}) -
\delta_{s,1} \fs \hstar \log \left(\frac{1+1/Y_{-}}{1+1/Y_{+}}\right),\\
\log Y_{a|uw} &= I_{a,a^{\prime}} \fs \star \log(1+Y_{a^{\prime}|uw}),
\end{align}
subjected to the asymptotic conditions
\begin{align}
\lim_{a\to \infty} \log Y_{a|uw}(\mu) = -2\mu\,a,
\qquad \lim_{s\to \infty}\log Y_{s|w}(h) = 2h\,s.
\end{align}

By furthermore performing the particle-hole transformations for all the particles assigned to the vertical wing of the fat hook,
that is $Y_{-}\to Y^{-1}_{-}$ and $Y_{a|uw}\to Y^{-1}_{a|uw}$, and making the following identifications,
$Y_{1,s+1}\equiv Y_{s|w}$ for $s\geq 1$, $Y_{a+1,1}\equiv Y_{a|uw}$ for $a\geq 1$ and $Y_{-}=Y_{1,1}$, $Y_{+}=Y_{2,2}$,
we recover the standard (universal) form of the $Y$-system functional relations. This time however,
unlike in the $\mathfrak{su}(N|M)$ chains, the corner node is just a different branch of the same electronic excitations and there
is no extra $U(1)$ chemical potential besides the charge and spin chemical potential $\mu$ and $h$, respectively.

The high-temperature $Y$-functions read explicitly
\begin{equation}
Y^{(0)}_{-}(\mu,h) = Y^{(0)}_{+}(\mu,h) = \frac{e^{\mu}+e^{-\mu}}{e^{h}+e^{-h}},\qquad
Y^{(0)}_{s|w}(h) = \chi^{2}_{s}(h)-1,\qquad
Y^{(0)}_{a|uw}(\mu) = \chi_{a}^{2}(\mu)-1,
\end{equation}
where, similarly as in the Heisenberg XXX model, the characters associated to the spin and charge wings are
\begin{equation}
\chi_{s}(h) = \frac{e^{-(s+1)h}-e^{(s+1)h}}{e^{-h}-e^{h}},\qquad
\chi_{a}(\mu) = \frac{e^{-(a+1)\mu}-e^{(a+1)\mu}}{e^{-\mu}-e^{\mu}}.
\end{equation}
Notice that functions $Y_{s|w}$ do not depend on $\mu$ and, likewise, $Y_{a|uw}$ do not depend on $h$. This means that
particles carrying charge are spin-less and, conversely, the ones carrying spin are charge-less.
It is only the `unbound electrons' which is charged under both degrees of freedom. At half filling, these simplify to
\begin{equation}
\lim_{\mu\to 0}Y^{(0)}_{a|uw}(\mu,h) = a(a+1),\qquad
\lim_{h\to 0}Y^{(0)}_{s|w}(\mu,h) = s(s+1),\qquad
\lim_{\mu=h\to 0}Y^{(0)}_{y}(\mu,h) = 1.
\end{equation}
The dressing transformation is written as a coupled system linear integral equations
\begin{align}
F_{s|w} - \fs \star I_{s,s^{\prime}}\ol{\vartheta}^{(0)}_{s^{\prime}|w}F_{s^{\prime}|w} +
\delta_{s,1}(\vartheta^{(0)}_{-}F_{-}-\vartheta^{(0)}_{+}F_{+}) &= 0,\\
F_{a|uw} - \fs \star I_{a,a^{\prime}}\ol{\vartheta}^{(0)}_{a^{\prime}|uw}F_{a^{\prime}|uw} -
\delta_{s,1}(\ol{\vartheta}^{(0)}_{-}F_{-}-\ol{\vartheta}^{(0)}_{+}F_{+}) &= 0,\\
F_{\pm} + \fs\star (\ol{\vartheta}^{(0)}_{1|w}F_{1|w} - \ol{\vartheta}^{(0)}_{1|uw}F_{1|uw}) &=
f^{\prime}_{\pm}-\fs \star f^{\prime}_{1|uw},
\label{eqn:Hubbard_dressing_transform}
\end{align}
where $\vartheta^{(0)}_{\pm} = \ol{\vartheta}^{(0)}_{\pm}=\tfrac{1}{2}$.
For the dressing of momentum (energy) we choose $f=k$ ($f=e$), with
\begin{align}
k^{\prime}_{\pm} &= \mp (1-u^{2})^{-1/2},\quad u\in [-1,1],\\
k^{\prime}_{1|uw}(u) &= k^{\prime}_{+}(u+\ii \uu)+k^{\prime}_{-}(u-\ii \uu),\quad u\in \mathbb{R}.
\end{align}

In the high-temperature limit, the dressed values of particles' spins $m^{\rm dr(0)}_{s|w}(h,\mu)$ and
charges $n^{\rm dr(0)}_{a|uw}(h,\mu)$ are computed from the logarithmic derivatives of the $Y$-functions,
\begin{align}
m^{\rm dr(0)}_{s|w}(h) &= \partial_{2h}\log Y^{(0)}_{s|w}(h) =
\frac{\partial_{h}\chi_{s}(h)}{\chi_{s}(h)-1/\chi_{s}(h)},&\qquad
m^{\rm dr (0)}_{y}(h) &= \partial_{2h}\log Y^{(0)}_{y}(h,\mu),\\
n^{\rm dr(0)}_{a|uw}(\mu) &= \partial_{2\mu}\log Y^{(0)}_{a|uw}(\mu) =
\frac{\partial_{\mu}\chi_{a}(\mu)}{\chi_{a}(\mu)-1/\chi_{a}(\mu)},&\qquad
n^{\rm dr (0)}_{y}(\mu) &= \partial_{2\mu}\log Y^{(0)}_{y}(h,\mu).
\end{align}
Notice also $m^{\rm dr (0)}_{a|uw} = 0$ and $n^{\rm dr (0)}_{s|w} = 0$.
An alternative route to compute the non-vanishing dressed spin and charge is to solve the following homogeneous
dressing transformation,
\begin{align}
C^{(0)}_{s,s^{\prime}}\star m^{\rm dr (0)}_{s^{\prime}|w} = 0,\qquad \lim_{s\to \infty}m^{\rm dr (0)}_{s|w} = 2s,\\
m^{\rm dr (0)}_{y} - \tfrac{1}{2}(\ol{\vartheta}^{(0)}_{1|uw}m^{\rm dr (0)}_{1|uw} - \ol{\vartheta}^{(0)}_{1|w}m^{\rm dr (0)}_{1|w}) = 0,
\end{align}
and similarly
\begin{align}
C^{(0)}_{a,a^{\prime}}\star n^{\rm dr (0)}_{a^{\prime}|uw} = 0,\qquad \lim_{a\to \infty}n^{\rm dr (0)}_{a|uw} = 2a,\\
n^{\rm dr (0)}_{y} - \tfrac{1}{2}(\ol{\vartheta}^{(0)}_{1|uw}n^{\rm dr (0)}_{1|uw} - \ol{\vartheta}^{(0)}_{1|w}n^{\rm dr (0)}_{1|w}) = 0,
\end{align}
whence we conclude
\begin{equation}
n^{\rm dr (0)}_{\pm}=\tfrac{1}{2}\vartheta^{(0)}_{1|uw}n^{\rm dr (0)}_{1|uw},\quad
m^{\rm dr (0)}_{\pm}=-\tfrac{1}{2}\vartheta^{(0)}_{1|uw}m^{\rm dr (0)}_{1|uw}.
\end{equation}
In particular, in the vicinity of the half-filled charge and spin sectors $h=0$ and $\mu=0$ we find
\begin{align}
m^{\rm dr (0)}_{s|w}(h) &\sim \frac{1}{3}(s+1)^{2}h + \mathcal{O}(h^{3}),\qquad
m^{\rm dr (0)}_{y}(h) \sim -\frac{1}{2}h + \mathcal{O}(h^{3}),\\
n^{\rm dr (0)}_{s|w}(\mu) &\sim \frac{1}{3}(a+1)^{2}\mu + \mathcal{O}(\mu^{3}),\qquad
n^{\rm dr (0)}_{y}(\mu) \sim \frac{1}{2}\mu + \mathcal{O}(\mu^{3}).
\end{align}

The derivatives of the dressed energies and momenta are computed from Eqs.~\eqref{eqn:Hubbard_dressing_transform}.
Indeed, the structure of the recurrence relations in the spin and charge wings of the fat hook take same form
as in the previously studied isotropic Heisenberg model, from where we readily obtain
the expressions for the $s|w$-strings and the $a|uw$-stacks
\begin{equation}
F^{(0)}_{a|uw} = \frac{a+1}{4}\left(\frac{f_{a|uw}}{a}-\frac{f_{a+2|uw}}{a+2}\right),\qquad
F^{(0)}_{s|w} = -\frac{s+1}{4}\left(\frac{f_{s|w}}{s}-\frac{f_{s+2|w}}{s+2}\right).
\end{equation}
Functions $F^{(0)}_{A}$ are interpreted as the derivatives of the dressed dressed momenta $p^{\prime (0)}_{A}$ or the
derivatives of the dressed energy $\varepsilon^{\prime (0)}_{A}$, depending whether the sources are chosen as
$f^{\prime}_{A} \leftarrow k^{\prime}_{A}$ or $f^{\prime}_{A} \leftarrow e^{\prime}_{A}$, respectively.
Taking into account that
$\ol{\vartheta}^{(0)}_{1|uw}f^{\prime (0)}_{1|uw}-\ol{\vartheta}^{(0)}_{1|w}f^{\prime (0)}_{1|w}=\tfrac{3}{2}f^{\prime (0)}_{1|uw}$, 
the remaining equation for the $y$-particles simplifies to
\begin{equation}
F^{\prime (0)}_{\pm}-\fs \star \tfrac{3}{2}F^{\prime (0)}_{1|uw} = f^{\prime}_{\pm} - \fs \star f^{\prime}_{1|uw}.
\end{equation}
Taking the sum and the difference and, using $f^{\prime}_{+}=-f^{\prime}_{-}$, we find
\begin{equation}
F^{\prime (0)}_{+} - F^{\prime (0)}_{+} = f^{\prime}_{+} - f^{\prime}_{-},\qquad
F^{\prime (0)}_{+} + F^{\prime (0)}_{+} = f^{\prime}_{2|uw}.
\end{equation}

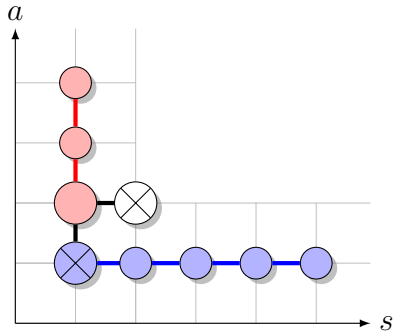
\begin{figure}[bh]
\centering

\begin{tikzpicture}[scale = 0.8]

\tikzstyle{node1}   = [circle, minimum width=16pt, draw, fill=blue!30!white, inner sep=0pt, path picture={\draw (path picture bounding box.south east) -- (path picture bounding box.north west) (path picture bounding box.south west) -- (path picture bounding box.north east);}, drop shadow]
\tikzstyle{node2}  = [circle, minimum width=16pt, draw, fill=red!30!white, inner sep=0pt, drop shadow]
\tikzstyle{node3}   = [circle, minimum width=16pt, draw, fill=white, inner sep=0pt, path picture={\draw (path picture bounding box.south east) -- (path picture bounding box.north west) (path picture bounding box.south west) -- (path picture bounding box.north east);}, drop shadow]
\tikzstyle{empty}  = [circle, minimum width=12pt, fill=white, draw, inner sep=0pt, drop shadow]
\tikzstyle{vert} = [circle, minimum width=12pt, fill = red!30!white, draw, drop shadow]
\tikzstyle{hor} = [circle, minimum width=12pt, fill = blue!30!white, draw, drop shadow]

\draw[very thin,color=lightgray] (0,0) grid (5.9,2);
\draw[very thin,color=lightgray] (0,0) grid (2,4.9);
\draw[->, >=latex, color=black] (0,0) -- (5.9,0) node[right] {\large $s$};
\draw[->, >=latex, color=black] (0,0) -- (0,4.9) node[above] {\large $a$};

\node[node1] (f1)  at (1,1) {};
\node[node2] (b)  at (1,2) {};
\node[node3] (f2) at (2,2) {};
\draw[-, ultra thick, color=blue] (f1) -- (2,1);
\foreach \x[count=\xnext from 1] in {1,...,4} {
		\node[hor] (s\x) at (\x+1,1) {};
	}
\foreach \x[count=\xnext from 2] in {1,...,3} {
	\draw[-, ultra thick, color=blue] (s\x) -- (s\xnext);
}	
\foreach \x[count=\xnext from 1] in {1,...,2} {
		\node[vert] (a\x) at (1,\x+2) {};
	}
	\foreach \x[count=\xnext from 2] in {1,...,1} {
	\draw[-, ultra thick, color=red] (a\x) -- (a\xnext);
}
\draw[-, ultra thick, color=red] (b) -- (a1);
\draw[-, ultra thick, color=black] (f1) -- (b);
\draw[-, ultra thick, color=black] (b) -- (f2);

\end{tikzpicture}
\caption{Thermodynamic particle content for (a) the Fermi--Hubbard model and (b) the $SU(2|2)$-symmetric chain of fundamental
particles with respect to the non-distinguished vacuum $\bigotimes\!\!-\!\!\bigodot\!\!-\!\!\bigotimes$. For (a), $(1,1)$ and $(2,2)$ 
are associated with the two-branched $y$-particle, while in (b) these separate into two distinct excitations which are
ascribed the $z_{\pm}$-roots.
Particles in the horizontal wing, sitting on nodes $(1,s+1)$ with $s\in \mathbb{N}$, are $s$-strings of $w$-roots for both (a)
and (b). Particles in the vertical wing, sitting on nodes $(a+1,1)$ with $a\in \mathbb{N}$, are (a) $uw$-stacks composed of both
the $y$-roots and $w$-roots or in (b) $z_{+}wz_{-}$-stacks made of $a+1$ $z_{+}$-roots, $a$ $w$-roots and $a-1$ $z_{-}$-roots.}
\end{figure}

\subsubsection*{$SU(2|2)$ integrable fermionic chain}

The $\mathfrak{su}(2|2)$-invariant model, also known as the EKS model introduced in \cite{EKS92}, is arguably the simplest
interacting integrable model of spin-full fermions on a one-dimensional lattice.
While the model exhibits certain structural similarities to the Fermi--Hubbard model, there are some important differences
to notice. Both models in fact arise as certain degenerate limits of a more general integrable model of spin-full lattice fermions 
with correlated hopping called the Hubbard--Shastry model~\cite{FQ12}.
It will be thus convenient to characterize the spectrum with respect to the non-distinguished vacuum, corresponding to
the following grading of local Hilbert space configurations, $|1|=|4|=0$ and $|2|=|3|=0$,
depicted by the following the Kac--Dynkin diagram
\begin{equation}
\bigotimes\!\!-\!\!\!-\!\!\bigodot\!\!-\!\!\!-\!\!\bigotimes:\qquad
\mathcal{K} = \begin{pmatrix}
0 & -1 & 0 \\
-1 & 2 & -1 \\
0 & -1 & 0
\end{pmatrix}.
\label{eqn:Cartan_matrix_2|2}
\end{equation}
The Bethe roots assigned to the nodes of the Kac--Dynkin diagram are
labelled by $z_{+,j}$, $w_{j}$ and $z_{-,j}$ when moving from the left to the right.

A short remark on the notation is in order here. The EKS model can be obtained from the Fermi--Hubbard model in the
scaling limit of large Coulomb repulsion $\uu \to 0$, causing the rapidity domain of the two-branched $y$-particle which lies along 
the branch cut $[-1,1]$ opening up to the whole real line and splitting into two distinct particles species whose roots
are denoted by $z_{+}$ and $z_{-}$. Notice that such a limiting procedure requires simultaneous rescaling the rapidity variable
to $u/\uu$ which recovers the standard parametrization of the elementary scattering kernels $K_{AB}(\uu)$
from Eq.~\eqref{eqn:kernels_def}. Likewise, the model can be understood as a weak-coupling limit of a more general
Hubbard--Shastry models \cite{FQ12}.

The bare energies and momenta of fermionic excitations become~\cite{FQ12}
\begin{align}
e_{+}(u) &= -2\cos{p_{+}(u)}-2 = -4\pi\,K_{1},&\qquad
e_{-}(u) &= -2\cos{p_{-}(u)}-2 = 0,\\
p_{+}(z_{+}) &= \frac{z_{+}+\ii}{z_{+}-\ii},&\qquad
p_{-} &= \pi.
\end{align}
This signifies that $z_{+}$ are the only momentum-carrying roots while $z_{-}$ cease to be dynamical.

As a consequence of the decoupling of the $y$-roots, the Bethe Ansatz equations now involve two nested levels
(assuming $L$ to be even)
\begin{align}
e^{\ii p_{+,k}L} &= \prod_{j=1}^{N_{w}}S^{-1}_{1}\left(z_{+,k},w_{j}\right),\\
-1 &= \prod_{j=1}^{N_{w}}S^{-1}_{2}\left(w_{k},w_{j}\right)\prod_{\alpha\in \pm}
\prod_{j=1}^{N_{\pm}}S_{1}\left(w_{k},z_{\pm,j}\right), \\
e^{\ii p_{-,k}L} &= (-1)^{L} = \prod_{j=1}^{N_{-}}S^{-1}_{1}(z_{-,k},w_{j}),
\end{align}
compatible with the fact that the rank of $\mathfrak{su}(2|2)$ equals three. The
scattering amplitudes are of the standard rational form, cf. Eq.~\eqref{eqn:fused_scattering_amplitudes}.
Eigenstates are uniquely parametrized in terms of $N_{\pm}$ roots of type $z_{\pm}$ and $N_{w}$ $w$-roots.
The $z_{+}$-roots are rapidities parametrizing bare momenta of electrons which occupy empty lattice sites.
A key difference with respect to the Fermi--Hubbard model is that instead of a single conservation of electrons $N_{y}$
we have two independent conserved $U(1)$ charges $\hat{N}_{+}$ and $\hat{N}_{-}$. The third conservation law is indeed
the Hubbard interaction $\hat{V}^{\rm H}$ which corresponds to conservation of doubly-occupied sites.
Notice also that to obtain configurations with doubly-occupied sites, all three types of roots need to be combined: one
begins by exciting singly occupied sites by adding $z_{+}$ $z_{-}$-roots, then by adding $w$-roots for a subset of them
one can lower their spin, and finally, a subset of sites with spin-down electrons one can add $z_{+}$-roots to add
the spin-up electrons.

The thermodynamic particle content of the $\mathfrak{su}(2|2)$ chain with respect to
the non-distinguished vacuum state $\bigotimes\!\!-\!\!\!-\!\!\bigodot\!\!-\!\!\!-\!\!\bigotimes$ consists of
\begin{itemize}
\item the $a|z_{+}wz_{-}$-stacks (with $a=1,2,\ldots$) representing bound state compounds made of
$a+1$ $z_{+}$ roots, $a$ $w$-roots and $a-1$ $z_{-}$ roots,
\item the auxiliary non-dynamical $s|w$-strings ($s=1,2,\ldots$) representing bound state of $s$ second-level $w$-roots,
with densities $\rho_{s|uw}$,
\item the $z_{\pm}$ roots, corresponding to two independent unbound fermionic excitations,
with $z_{+}$ being corresponding to physical momentum-carrying electronic excitations and $z_{-}$ being the third-level
auxiliary real Bethe roots.
\end{itemize}
The rapidity derivatives of the bare momenta and energies for the momentum-carrying particles are
\begin{align}
p^{\prime}_{+}(u) &= -2\pi\,K_{1}(u),&\qquad
e_{+}(u) &= -4\pi\,K_{1}(u),\\
p^{\prime}_{a|z_{+}wz_{-}}(u) &= -2\pi\,K_{a+1}(u),&\qquad
e_{a|z_{+}wz_{-}}(u) &= -4\pi\,K_{a+1}(u)
\end{align}
For the non-dynamical particles we obviously have $p^{\prime}_{s|w}=0$ and $p^{\prime}_{-}=0$.

The particles are charged under three Cartan charges, involving two independent bosonic $U(1)$ global charges 
$\hat{S}^{z}$, $\hat{\eta}^{z}$ (coupling to chemical potentials $2h$ and $2\mu$, respectively) associated with the spin and charge 
global generators of the two distinct $\mathfrak{su}(2)$ sectors, and an additional fermionic $U(1)$ charge $\hat{V}^{\rm H}$
which couples to chemical potential $\uu$ (contributing a constant shift to particles' bare dispersions).
The values of bare charges are
\begin{align}
m_{a|z_{+}wz_{-}} &= 0,\quad m_{s|w} = -1,\quad m_{\pm} = \tfrac{1}{2},\\
n_{a|z_{+}wz_{-}} &= 1,\quad n_{s|w} = 0,\quad n_{\pm} = \tfrac{1}{2},\\
n^{\rm H}_{a|z_{+}wz_{-}} &= 1,\quad n^{\rm H}_{s|w} = 0,\quad n^{\rm H}_{\pm} = \mp \tfrac{1}{2}.
\end{align}
The high-temperature limit of the free energy density, $\lim_{\beta \to 0}f_{2|2}(\beta;h,\mu,\uu)$, thus reads
\begin{equation}
f^{(0)}_{2|2}(h,\mu,\uu) = \sum_{\alpha \in \pm}(-h-\mu \mp 2\uu)\star \rho_{\pm}
+ (-2a\mu-4\uu)\star \rho_{a|z_{+}wz_{-}} + 2h\,s\star \rho_{s|w}.
\end{equation}

The quasi-local TBA equations read
\begin{align}
\log Y_{\pm} &= \nu_{\pm} + \fs\star \log\frac{1+Y_{1|uw}}{1+Y_{1|w}},\\
\log Y_{a|z_{+}wz_{-}} &= \nu_{a|uw} + I_{a,a^{\prime}}\fs\star \log(1+Y_{a^{\prime}|uw})
+ \delta_{a,1}\fs \star \log\frac{1+Y_{+}}{1+Y_{-}},\\
\log Y_{s|w} &= \nu_{s|w} + I_{s,s^{\prime}}\fs\star \log(1+Y_{s^{\prime}|uw})
+ \delta_{s,1}\fs \star \log\frac{1+1/Y_{+}}{1+1/Y_{-}},
\label{eqn:local_TBA_EKS}
\end{align}
with source terms
\begin{equation}
\nu_{+}(u) = \beta \left(e_{+}(u) - \fs \star e_{1|uw}(u)\right),\qquad
\nu_{-}(u) = \beta \left(e_{-}(u) - \fs \star e_{1|uw}(u)\right) + 4\uu,\qquad
\nu_{a|z_{+}wz_{-}} = \nu_{s|w} = 0,
\label{eqn:local_source_EKS}
\end{equation}
and asymptotics
\begin{equation}
\lim_{a\to \infty} \log Y_{a|z_{+}wz_{-}} = -2\mu\,a,\qquad
\lim_{s\to \infty} \log Y_{s|w} = 2h\,s.
\label{eqn:asymptotics_EKS}
\end{equation}
Let us stress that parameter $\uu$, associated to the Hubbard charge $\hat{V}^{\rm H}$, does not enter via the asymptotics,
but instead explicitly appears in the equation for the distinguished corner node $(2,2)$.
Equations \eqref{eqn:local_source_EKS} take the standard $Y$-system format upon performing subtable particle-hole transformations along the vertical wing of the fat hook, namely
\begin{equation}
Y_{1,1} \equiv Y^{-1}_{+},\quad Y_{2,2} \equiv Y_{-},\quad Y_{a+1,1}\equiv Y^{-1}_{a|uw},\quad Y_{1,s+1}\equiv Y_{s|w}.
\end{equation}
In the high-temperature limit the quasi-local TBA equations take the form of coupled algebraic equations
\begin{align}
\left[Y^{(0)}_{+}\right]^{2} &= \left[e^{-4\uu}Y^{(0)}_{-}\right]^{2} = \frac{1+Y^{(0)}_{1|uw}}{1+Y^{(0)}_{1|w}},\\
\left[Y^{(0)}_{a|z_{+}wz_{-}}\right]^{2} &= \left(1+Y^{(0)}_{a-1|uw}\right)\left(1+Y^{(0)}_{a+1|uw}\right)
\left(\frac{1+1/Y^{(0)}_{+}}{1+1/Y^{(0)}_{-}}\right)^{\delta_{a,1}},\\
\left[Y^{(0)}_{s|w}\right]^{2} &= (1+Y^{(0)}_{s-1|w})(1+Y^{(0)}_{s+1|w})
\left(\frac{1+Y^{(0)}_{+}}{1+Y^{(0)}_{-}}\right)^{\delta_{s,1}}.
\end{align}
The solution to these equations will once again be given in terms of $\chi$-functions.
For the horizontal (spin) and vertical (charge) wings we find
\begin{align}
\chi_{1,s\geq 2}(h,\mu,\uu) &= \left(e^{\mu}+e^{-\mu}\right)\frac{\sinh{(s\,h)}}{\sinh{h}}+
e^{-2\uu}\frac{\sinh{((s-1)\,h)}}{\sinh{h}}+e^{2\uu}\frac{\sinh{((s+1)\,h)}}{\sinh{h}},\\
\chi_{a\geq 2,1}(h,\mu,\uu) &= T_{1,a}(\mu,h,-\uu).
\end{align}
We have adopted a convenient symmetric gauge-fixing condition, given by $\chi_{0,s}=\chi_{a,0}=1$ ($a\geq 0$ and $s\in \mathbb{Z}$)
for the exterior boundary, and
\begin{equation}
\chi_{2,s\geq 3}(h,\mu,\uu) = \chi_{a\geq 3,2}(h,\mu,\uu) =
4\left(2\cosh{(h)}\cosh{(\mu)}\cosh{(2\uu)}+\cosh^2{(h)}+\cosh^{2}{(\mu)}+\sinh^{2}{(2\uu)}\right),
\end{equation}
for the $\chi$-functions assigned to the interior boundary of the fat hook.
Such a gauge choice render the particle-hole symmetry between the spin and charge wings manifest, but it comes with a price
since it force us to define two independent $\chi$-functions at the corner node
\begin{equation}
\chi^{\rm \rightarrow}_{2,2}(h,\mu,\uu) = \chi^{\rm \uparrow}_{2,2}(\mu,h,-\uu).
\end{equation}
This is permissible as all the $\chi$-functions are uniquely and unambiguously fixed by the requirement that
the classical Hirota bilinear relations hold in the respective wings. Nothing in principle prevents us defining
a unique corner $\chi$-function, but doing this seems less natural as it generates asymmetry between the wings.
At any rate, it is the $Y$-functions which are gauge-invariant object and thus have a physical meaning.

The character at the fundamental node is the logarithm of the free energy density
$f^{(0)}_{2|2}=-\log \chi_{1,1}(h,\mu,\uu)$, with
\begin{equation}
\chi_{1,1}(h,\mu,\uu) = e^{\uu}(e^{h}+e^{-h}) + e^{-\uu}(e^{\mu}+e^{-\mu}).
\end{equation}

The dressing transformation in the high-temperature limit has the same structure as previously in the Hubbard model.
In the present convention, it reads explicitly
\begin{align}
F_{\pm} - \fs \star \left(\ol{\vartheta}^{(0)}_{1|z_{+}wz_{-}}F_{1|z_{+}wz_{-}}-\ol{\vartheta}^{(0)}_{1|w}F_{1|w}\right) &=
f_{\pm} - \fs \star f_{1|z_{+}w z_{-}},\\
F_{s|w} - \fs \star I_{s,s^{\prime}}\ol{\vartheta}^{(0)}_{s^{\prime}|w}F_{s^{\prime}|w} -
\delta_{s,1} \left(\vartheta^{(0)}_{-}F_{-} - \vartheta^{(0)}_{+}F_{+}\right) &=0,\\
F_{s|z_{+}w z_{-}} - \fs \star I_{s,s^{\prime}}\ol{\vartheta}^{(0)}_{s^{\prime}|z_{+}w z_{-}}F_{s^{\prime}|z_{+}w z_{-}} +
\delta_{a,1} \left(\ol{\vartheta}^{(0)}_{-}F_{-} - \ol{\vartheta}^{(0)}_{+}F_{+}\right) &=0.
\end{align}
Recall that there is an implicit dependence on all three $U(1)$ chemical potentials entering via the mode occupation functions.

As usual, we now inspect the properties of the dressed $U(1)$ charges in the high-temperature limit.
The dressed spin and charge in the vicinity of the half-filled spin and charge sector respectively read,
\begin{equation}
m^{\rm dr (0)}_{s|w}(h,\mu,\uu) \sim \zeta^{(m)}_{s}(\mu,\uu)\,h + \mathcal{O}(h^{3}),\qquad
m^{\rm dr (0)}_{a|uw}(h,\mu,\uu) \sim \zeta^{(m)}_{a}(\mu,\uu)\,h + \mathcal{O}(h^{3}).
\end{equation}
and
\begin{equation}
n^{\rm dr (0)}_{s|w}(h,\mu,\uu) \sim \zeta^{(n)}_{s}(h,\uu)\,\mu + \mathcal{O}(\mu^{3}),\qquad
n^{\rm dr (0)}_{a|uw}(h,\mu,\uu) \sim \zeta^{(n)}_{a}(h,\uu)\,\mu + \mathcal{O}(\mu^{3}).
\end{equation}

One key difference compare to the Hubbard model worth pointing out  is that now the dressed spin (resp. charge) for
any finite values of $\uu$ in the half-filled spin (resp. charge) sector depends explicitly on the
other two chemical potentials, namely $\mu$ (resp. $h$) and $\uu$. This can be understood from the fact that
the third-level (non-dynamical) Bethe roots $u^{(3)}_{j}\equiv z_{-,j}$ participate in the formation of doubly occupied lattice sites.
In fact, since finite $\uu$ induces imbalance between $S^{z}$-spin and $\eta^{z}$-spin, the dressed spin and charges
satisfy the following symmetry relations
\begin{equation}
m^{\rm dr(0)}_{s|w}(h,\mu,\uu) = n^{\rm dr (0)}_{a|z_{+}wz_{-}}(\mu,h,-\uu),\qquad
m^{\rm dr(0)}_{a|z_{+}wz_{-}}(h,\mu,\uu) = n^{\rm dr (0)}_{s|w}(\mu,h,-\uu),
\end{equation}
upon interchanging spin with charge $s\leftrightarrow a$ and flipping the sign of $\uu$, $\uu \to -\uu$.
It is thus sufficient to examine the behaviour close the half-filled spin sector.
We are not interested in the most general solution but mostly
in the large-$s$ behavior. To this end it is useful to introduce the following ratios of the $\chi$-functions
in the horizontal and vertical wings, namely
\begin{equation}
g^{\rightarrow}_{s\geq 1} = \frac{\chi_{1,s+1}}{\chi_{0,s+1}\chi_{2,s+1}},\qquad
g^{\uparrow}_{a\geq 1} = \frac{\chi_{a+1,1}}{\chi_{a+1,0}\chi_{a+1,2}}.
\end{equation}
The high-temperature limit of the dressed spin read
\begin{equation}
m^{\rm dr (0)}_{s}(h,\mu,\uu) = \frac{\partial_{h}g_{s}(\mu,\uu)}{g_{s}(\mu,\uu)-1/g_{s}(\mu,\uu)}.
\end{equation}

\subsubsection*{$SU(2|1)$ spin chain (SUSY t--J model)}

The Hamiltonian of the t--J model expressed in terms of spin-full electrons takes the following form
\begin{equation}
\hat{H}_{\rm t-J} = \hat{P}\Big[-{\rm t}\sum_{j,\sigma}\hat{c}^{\dagger}_{j,\sigma}\hat{c}_{j+1,\sigma}+
\hat{c}^{\dagger}_{j+1,\sigma}\hat{c}_{j,\sigma}\Big]\hat{P}
+{\rm J}\sum_{j}\left(\hat{\vec{S}}_{j}\cdot \hat{\vec{S}}_{j+1}-\tfrac{1}{4}\hat{n}_{j}\hat{n}_{j+1}\right),
\label{eqn:tJ_Hamiltonian}
\end{equation}
where $\hat{P}=\prod_{j=1}^{L}(1-\hat{n}_{j\da}\hat{n}_{j\ua})$ has been used project out configurations with doubly occupied sites.
The model becomes integrable at the `supersymmetric point' ${\rm J}=2{\rm t}$, where $\hat{H}_{\rm t-J}$ becomes
proportional to the $SU(2|1)$-symmetric Hamiltonian $\hat{H}^{2|1}$.
The SUSY t--J model can also be retrieved from the large-repulsion $\uu \to \infty$ limit of the Hubbard model.
The model can be understood as an extension of the $\mathfrak{su}(2)$ Heisenberg chain by introducing vacant sites
and treat them as the electron holes (i.e. fermions).

We find it most convenient to formulate the problem in the distinguished grading
$|0|=|1|=0$ and $|2|=1$, corresponding to the diagram
\begin{equation}
\bigodot\!\!-\!\!\!-\!\!\bigotimes:\qquad
\mathcal{K} = \begin{pmatrix}
2 & -1 \\
-1 & 0
\end{pmatrix}.
\end{equation}
The vacuum state here is a completely polarized (ferromagnetic) state.
The primary (physical) excitations are momentum-carrying spin-down magnonic excitations which form bound states,
described by the primary Bethe roots $u^{(1)}_{j}$. The bare momentum of an elementary is
\begin{equation}
k(u^{(1)}_{j})=\ii \log S_{1}\left(u^{(1)}_{j}\right).
\end{equation}
In addition, we have an extra specie of fermionic excitations corresponding to vacancies (i.e. holes) of electrons,
described by the second-level (auxiliary) rapidities $u^{(2)}_{\alpha}$. The latter do not carry momenta and energy.
The Bethe equations with respect to the ferromagnetic background take the form
\begin{align}
e^{\ii k(u^{(1)}_{j})L}\prod_{k=1}^{N_{1}}S_{2}\left(u^{(1)}_{j},u^{(2)}_{k}\right)
\prod_{l=1}^{N_{2}}S^{-1}_{1}\left(u^{(1)}_{j},u^{(2)}_{l}\right) &= -1,\\
\prod_{j=1}^{N_{1}}S_{1}\left(u^{(2)}_{l},u^{(1)}_{j}\right) &= 1.
\end{align}
The number primary Bethe roots $u^{(1)}_{j}\in \mathbb{C}$
is $N_{1}$, while the number of real charge rapidities $u^{(2)}_{\alpha}$ is $N_{2}$.
The number of roots obey the following inequalities
\begin{equation}
N_{2} \leq N_{1},\qquad N_{1} \leq \frac{1}{2}(L+N_{2})\leq L.
\end{equation}
Here $N_{1}=N_{h}+N_{\da}$ is the number of hole plus spin-down excitations and $N_{2}=N_{h}=N_{1}-N_{\da}$ is the total
number of electron charge holes. The total spin and electron charge are
$S^{z}=\tfrac{1}{2}(N_{\ua}-N_{\da})=\tfrac{1}{2}(L-2N_{1}+N_{2})$ ($N_{\ua}=L-N_{h}-N_{\da}$) and
$N_{e}=N_{\ua}+N_{\da}$ respectively. The $U(1)$ chemical potentials which couple to total spin $\hat{S}^{z}$ and number of holes
$\hat{N}_{h}=1-\hat{N}_{e}$ are denoted by $2h$ and $\mu$, respectively. The electron filling fraction is
$\alpha_{\rm c} = N_{e}/(N_{e}+N_{h})=1-N_{h}/L$. The total energy of a state is the sum of all spin-down excitations 
$E \simeq \sum_{j=1}^{N_{1}}2\pi\,K_{1}(u^{(1)}_{j})$.

To reconcile the notation of the one used above, we relabel the Bethe roots as $u^{(1)}_{j}\to u_{j} \in \mathbb{C}$ and
$u^{(2)}_{l}\to w_{l} \in \mathbb{R}$. The former represent charge-less bound spin excitations carrying bare spin
\begin{equation}
m_{s}=-s,\quad n_{s}=0,
\end{equation}
which form the standard $s$-strings with real centres
\begin{equation}
u^{(s)}_{j,k} = u^{(s)}_{j} + \tfrac{\ii}{2}(s+1-2k),\quad k = 1\sim s,
\end{equation}
while the fermionic roots $w_{l}$ correspond to electron holes which do not form bound states.
The total number of primary Bethe roots is the number of spin-down excitations, $N_{1}=\sum_{s=1}^{\infty}s\,N_{s}$.
Adding an electron hole amounts to remove a spin-up electron excitation and hence
\begin{equation}
m_{\otimes}=-\tfrac{1}{2},\quad n_{\otimes}=-1.
\end{equation}

\begin{figure}
\centering

\begin{tikzpicture}[scale = 0.8]

\tikzstyle{bosonic}  = [circle, minimum width=16pt, draw, fill=blue!30!white, inner sep=0pt, drop shadow]
\tikzstyle{fermionic}   = [circle, minimum width=16pt, draw, fill=red!30!white, inner sep=0pt, path picture={\draw (path picture bounding box.south east) -- (path picture bounding box.north west) (path picture bounding box.south west) -- (path picture bounding box.north east);}, drop shadow]
\tikzstyle{empty}  = [circle, minimum width=12pt, fill=white, draw, inner sep=0pt, drop shadow]
\tikzstyle{vert} = [circle, minimum width=12pt, fill = red!30!white, draw, drop shadow]
\tikzstyle{hor} = [circle, minimum width=12pt, fill = blue!30!white, draw, drop shadow]

\draw[very thin,color=lightgray] (0,0) grid (5.9,2);
\draw[very thin,color=lightgray] (0,0) grid (1,3.9);
\draw[->, >=latex, color=black] (0,0) -- (5.9,0) node[right] {\large $s$};
\draw[->, >=latex, color=black] (0,0) -- (0,3.9) node[above] {\large $a$};

\node[bosonic] (b)  at (1,1) {};
\node[fermionic] (f) at (1,2) {};
\draw[-, ultra thick, color=blue] (b) -- (2,1);
\foreach \x[count=\xnext from 1] in {1,...,4} {
		\node[hor] (s\x) at (\x+1,1) {};
	}
\foreach \x[count=\xnext from 2] in {1,...,3} {
	\draw[-, ultra thick, color=blue] (s\x) -- (s\xnext);
}	
\draw[-, ultra thick, color=black] (b) -- (f);

\end{tikzpicture}
\caption{Thermodynamic particle content for the $SU(2|1)$ fundamental spin chain (SUSY integrable t--J model) with respect to
the distinguished bare vacuum $\bigodot\!\!-\!\!\!-\!\!\bigotimes$. The momentum-carrying particles are bosonic $s$-strings
attached to nodes $(1,s)$, $s \in \mathbb{N}$. The corner node $(2,1)$ is assigned an auxiliary fermionic excitation
representing electron vacancies.}
\end{figure}
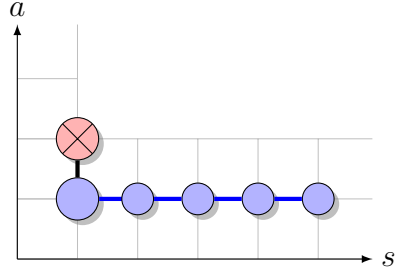

Introducing rapidity distributions $\rho_{\otimes}$ and $\rho_{s}$, the canonical TBA equations take the form
\begin{align}
\log Y_{s} &= \mu_{s} - K_{s}\star \log(1+Y^{-1}_{\otimes}) +  K_{s,s^{\prime}}\star \log(1+Y^{-1}_{s^{\prime}}),\\
\log Y_{\otimes} &= \mu_{\otimes} - K_{s}\star \log(1+Y^{-1}_{s}),
\label{eqn:canonical_TBA_2|1_distinguished}
\end{align}
with chemical potentials
\begin{equation}
\mu_{\circ|s} =-\beta\,e_{\circ|s} + 2h\,s,\qquad \mu_{\otimes} = \mu+h,
\end{equation}
where $e_{s}=2\pi\,K_{s}$ are the bare energies of $s$-strings.

It is instructive to remark that Eqs.~\eqref{eqn:canonical_TBA_2|1_distinguished} are just a particular singular reduction
of the $\mathfrak{su}(2|2)$ canonical TBA equations written for e.g. the distinguished
(ferromagnetic) vacuum $\bigodot\!\!-\!\!\!-\!\!\bigotimes\!\!-\!\!\!-\!\!\bigodot$.
Such a reduction is realized by freezing appropriate charge degrees of freedom in order to prohibit double occupancies while still 
allowing empty sites. Specifically, this amounts to remove the third-level Bethe roots responsible for the doubly-occupied 
configurations, accompanied by decoupling the following subset of the $Y$-functions, $Y_{a,1}\to 0$, for $a\geq 3$, and
$Y_{2,2}\to \infty$.
\\

Below we outline how to transform the canonical TBA equations for the $\mathfrak{su}(2|1)$ chain to the quasi-local form.
The starting point are the quasi-local TBA equations of the $\mathfrak{su}(2|2)$ chain, which in the limit described above
become
\begin{align}
\log Y_{1,s} &= \nu_{1,s} - \delta_{s,1}\fs \star \log(1+1/Y_{2,1})
+ I_{s,s^{\prime}} \fs \star \log(1+Y_{1,s^{\prime}}), \\
\log Y_{2,1} &= \nu_{2,1} - K_{1}\star \log(1+Y_{1,1})
- K_{2}\star \log(1+Y_{2,1}),
\end{align}
supplemented with large-$s$ asymptotics
\begin{equation}
\lim_{s\to \infty}\log Y_{1,s}(h,\mu) = 2h\,s.
\end{equation}
These equations are equivalent to those presented previously in ref. \cite{Frahm99}, which can be readily confirmed
by first convolving with respect to $\fs$ and subsequently deconvolving with respect to $K_{1}$,
\begin{align}
\log Y_{s} &= \nu_{s} + I_{s,s^{\prime}}\fs \star \log(1+Y_{s^{\prime}})
-\delta_{s,1}\fs\star \log(1+1/Y_{\otimes}),\\
\log Y_{\otimes} &= \nu_{\otimes} -\fs \star \log(1+Y_{1})
- \widetilde{K}\star \log(1+1/Y_{\otimes}),
\label{eqn:local_TBA_2|1_distinguished}
\end{align}
with $\hat{\widetilde{K}}(\kappa)=e^{-|k|}(1+e^{-|k|})^{-1}$. The source terms are of the form
\begin{equation}
\nu_{s} = -\beta\,\delta_{s,1}\fs,\qquad
\nu_{\otimes} = -\beta(\fs \star e_{1}) + \mu.
\end{equation}
The high-temperature dressing transformation takes the form
\begin{align}
F_{s} - \fs \star I_{s,s^{\prime}}\ol{\vartheta}^{(0)}_{s^{\prime}}F_{s^{\prime}} -
\delta_{s,1}\fs \star \vartheta^{(0)}_{\otimes}F_{\otimes} &= -\delta_{s,1}\fs,\\
F_{\otimes} + \fs \star \ol{\vartheta}^{(0)}_{1}F_{1} - \widetilde{K}\star \vartheta^{(0)}_{\otimes}F_{\otimes} &= -\fs \star K_{1}.
\label{eqn:highT_dressing_2|1}
\end{align}

We proceed by analysing the dressed spin and charge and their dependence on the grand-canonical chemical potentials.
In the high-temperature limit $\beta\to 0$ the $Y$-functions become constant
and Eqs.~\eqref{eqn:local_TBA_2|1_distinguished} turns into a set of algebraic relations
\begin{align}
\left[Y^{(0)}_{s}\right]^{2} &= \frac{(1+Y^{(0)}_{s-1})(1+Y^{(0)}_{s+1})}{(1+1/Y^{(0)}_{\otimes})^{\delta_{s,1}}},\\
\left[e^{-\mu}Y^{(0)}_{\otimes}\right]^{2} &= \frac{1}{(1+1/Y^{(0)}_{\otimes})(1+Y^{(0)}_{1})}.
\label{eqn:Y21_local_TBA_distinguished}
\end{align}
These take the standard form of the $Y$-system functional relations by identifying $Y_{s}\equiv Y_{1,s}$ and
$Y_{\otimes}\equiv Y_{2,1}$. The self-coupling term in the last equation for the exceptional corner node is the
remainder of the collapsed vertical wing of the $\mathfrak{su}(2|2)$ spectrum which leaves
behind explicit dependence on the charge chemical potential $\mu$.
This is reminiscent of the spin chemical potential entering explicitly in the truncated spectrum of the gapless regime
of the XXZ Heisenberg model which was due to the root-of-unity restriction.

The solution to Eqs.~\eqref{eqn:Y21_local_TBA_distinguished} is
\begin{equation}
Y_{s}(h,\mu) = \left(\frac{e^{-(s+1)h+\Phi}-e^{(s+1)h+\Phi}}{e^{-h}-e^{h}}\right)^{2}-1,\qquad
Y_{\otimes}(h,\mu) = \frac{e^{-\mu}}{e^{-h}+e^{h}+e^{\mu}},
\end{equation}
with
\begin{equation}
\Phi(h,\mu) = \frac{1}{2}\log\left(\frac{1+e^{-h+\mu}}{1+e^{h+\mu}}\right).
\end{equation}

The high-temperature limit of the grand canonical free energy density $f^{(0)}_{2|1}=f^{(0)}_{2|1}(h,\mu)$ is defined as
\begin{equation}
f^{(0)}_{2|1}(h,\mu) = -\lim_{L\to \infty}\frac{1}{L}\log {\rm Tr}\,\exp{\left(2h\,\hat{S}^{z}+\mu\,\hat{N}_{h}\right)}.
\end{equation}
In terms of the $Y$-functions we have
\begin{equation}
f^{(0)}_{2|1}(h,\mu) = -\tfrac{1}{2}\log\left[\big(1+Y^{(0)}_{\circ|1}(h,\mu)\big)\big(1+1/Y^{(0)}_{\otimes}(h,\mu)\big)\right]
= -\log \chi_{1,1}(h,\mu),
\end{equation}
with the fundamental $\chi$-function reading
\begin{equation}
\chi^{(0)}_{1,1}(h,\mu) = e^{h}+e^{-h}+e^{\mu}.
\end{equation}
By imposing boundary conditions $\chi_{0,s}=1$ and $\chi_{a\geq 0,0}=1$, together with $\chi^{(0)}_{1,1}$ and $\chi^{(0)}_{2,1}$ 
determined from $1+Y^{(0)}_{1,1}=[\chi^{(0)}_{1,1}]^{2}/\chi^{(0)}_{2,1}$, we unique fix all $\chi^{(0)}_{a,s}$ on the
$(a,s)$-lattice. Specifically, the infinite tower of symmetric characters $\chi_{1,s}$ for $s\geq 1$ can be calculated 
recursively
\begin{equation}
\chi^{(0)}_{1,s+1}(h,\mu) = (e^{h}+e^{-h})\chi^{(0)}_{1,s}(h,\mu) - \chi^{(0)}_{1,s-1}(h,\mu).
\end{equation}

In the high-temperature limit, the dressed spin and charge are calculated as
\begin{equation}
m^{\rm dr(0)}_{s}(h,\mu) = \partial_{2h}\log Y^{(0)}_{s}(h,\mu),\qquad
n^{\rm dr(0)}_{s}(h,\mu) = \partial_{\mu}\log Y^{(0)}_{s}(h,\mu).
\end{equation}
The dressed values of particles' spin can likewise be obtained from the following rapidity-independent recurrence relation
\begin{equation}
m^{\rm dr(0)}_{s} - \tfrac{1}{2}I_{s,s^{\prime}}\ol{\vartheta}^{(0)}_{s^{\prime}}m^{\rm dr(0)}_{s^{\prime}} = 0,
\qquad \lim_{s\to \infty}m^{\rm dr(0)}_{s} = s.
\end{equation}
We subsequently specialize our attention to the half-filled spin sector.
In the vicinity of the half-filled spin sector $h=0$, $S^{z}(h,\mu)=\partial_{2h} f^{(0)}_{\rm gc}(h,\mu)$
(with $\lim_{h\to 0}S^{z}(h,\mu)=0$ and $\lim_{h\to \pm \infty}S^{z}(h,\mu)=\pm \tfrac{1}{2}$, irrespective of $\mu$) we have
\begin{align}
m^{\rm dr(0)}_{s}(h,\mu) & = \frac{h}{6}\left(\frac{6}{(e^{\mu}+1)^{2}}-\frac{6}{e^{\mu}+1}+2s^{2}+
\frac{s(2s-1)}{e^{\mu}(s+1)-s}+\frac{(s+2)(2s+3)}{e^{\mu}(s+1)+s+2}\right) + \mathcal{O}(h^{3}),\\
m^{\rm dr(0)}_{\otimes}(h,\mu) &= -\frac{h}{e^{\mu}+2} + + \mathcal{O}(h^{3}).
\end{align}
For $\mu = 0$ these expressions further simplify to
\begin{equation}
m^{\rm dr(0)}_{s}(h,0) \sim \tfrac{h}{12}(2s+1)^{2}+\mathcal{O}(h^{3}),\qquad
m^{\rm dr(0)}_{\otimes}(h,0) \sim -\tfrac{h}{3}+\mathcal{O}(h^{3}).
\end{equation}
The vanishing of the dressed spin at half filling (irrespectively of the chemical potential $\mu$) is actually implied by the 
bosonic symmetry, realized by performing the spin-reversal transformation on the bosonic states.
The electron charge transport behaves quite differently however. Let us examine the dressed charge in the high-temperature limit,
given by
\begin{equation}
n^{\rm dr(0)}_{s}(h,\mu) = \partial_{\mu} \log Y_{s}(h,\mu),\qquad
n^{\rm dr(0)}_{\otimes}(h,\mu) = \partial_{\mu} \log Y_{s}(h,\mu) = -\frac{e^{-h}+e^{h}+2e^{\mu}}{e^{-h}+e^{h}+e^{\mu}}.
\end{equation}
For instance, the hole excitations propagating in the half-filled spin background carry finite dressed charges
\begin{equation}
\lim_{h\to 0}n^{\rm dr(0)}_{s}(h,\mu) = -\frac{2e^{\mu}(s(e^{\mu}+1)+1)}{(e^{\mu+1})(e^{\mu}(s-1)+s)(e^{\mu}(s+1)+s+2)},\qquad
\lim_{h\to 0}n^{\rm dr(0)}_{\otimes}(h,\mu) = -\frac{2(e^{\mu}+1)}{e^{\mu}+2}.
\end{equation}
The spin degrees of freedom can be excited independently of hole excitations. The addition of a hole implies removing
a spin-up electron from the state. Notice moreover that $N_{h}(h,\mu)=-\partial_{\mu} f^{(0)}_{\rm 2|1}(h,\mu)$ and hence
the vanishing chemical potential corresponds to the third-filling $N_{h}(0,0)=\tfrac{1}{3}$
(for $\mu \to -\infty$ we have $N_{h}=0$).
Hence, imposing the filling fraction $\alpha_{\rm c}=\lim_{L\to \infty}N_{\rm c}/L$ ($0\leq \alpha_{\rm c}\leq 1$)
for arbitrary value of $h$ requires to adjust $\mu$ in accordance with
\begin{equation}
e^{\mu} = \frac{e^{-h}\alpha_{\rm c} + e^{h}\alpha_{\rm c}}{1-\alpha_{\rm c}}.
\end{equation}
In particular, for the half-filled charge sector $\alpha_{\rm c}=\tfrac{1}{2}$ this means $e^{\mu}=e^{-h}+e^{h}$.

The high-temperature mode occupation functions in the half-filled spin sector ($h=0$) and
the third-filled charge sector $\mu=0$ are
\begin{align}
\lim_{\mu \to 0}\lim_{h\to 0} \vartheta^{(0)}_{s}(h,\mu) &= \frac{4}{(2s+1)^{2}},\\
\lim_{\mu \to 0}\lim_{h\to 0} \vartheta^{(0)}_{\otimes}(h,\mu) &= \frac{3}{4}.
\end{align}
The $s$-string bound state corresponds to atypical (short) irreducible $\mathfrak{su}(2|1)$ representations which are
of dimension $d_{1,s}=\lim_{G\to 1}\chi_{1,s}(G)=2s+1$.
The solution to the dressing equation \eqref{eqn:highT_dressing_2|1} in the limit $\mu \to 0$ and $h\to 0$ reads
\begin{equation}
F_{s} = \frac{2s+1}{3}\left(\frac{K_{s}}{2s-1}-\frac{K_{s+2}}{2s+3}\right),\qquad
F_{\otimes} = \frac{4}{9}K_{2}.
\end{equation}
\\

The conclusion of above analysis is that for all finite values of the charge chemical potential $\mu$, the dressed 
charges $n^{\rm dr(0)}_{s}$ and $n^{\rm dr(0)}_{\otimes}$ always remain positive definite quantities.
Near the half-filled spin sector $h\to 0$  and the third-filled charge sector $\mu \to 0$, the dressed electron charges
read explicitly
\begin{align}
\lim_{h\to 0}n^{\rm dr(0)}_{s}(h,\mu) &\sim
\frac{2s+1}{(2s-1)(2s+3)} - \frac{4s^{2}+4s+5}{2(2s-1)^{2}(2s+3)^{2}} \mu + \mathcal{O}(\mu^{2}),\\
\lim_{h\to 0}n^{\rm dr(0)}_{\otimes}(h,\mu) &\sim \frac{4}{3} - \frac{2}{9}\mu + \mathcal{O}(\mu^{2}).
\end{align}

\section{Lower bound on diffusion constants}

In this section we re-derive a relation between the linear-response diffusion constants and the corresponding Drude weights,
originally presented ref.~\cite{PhysRevLett.119.080602}. We consider the linear transport the conserved $U(1)$ charges
\begin{equation}
\hat{Q} = \sum_{x=-L/2}^{L/2-1} \hat{q}_{x},
\end{equation}
Below we derive an explicit lower bound on the charge diffusion constant $D$.
For simplicity we first specialize to the infinite-temperature Gibbs equilibrium, and assume that the local charge density $q$ has
$d$ distinct eigenvalues $q\in \{-(d-1)/2,\dots,(d-1)/2\}$ of the same multiplicity.

We consider a lattice of length $L$, with an initial state described by the following density matrix
\begin{equation}
\hat{\varrho}(\beta,\delta h)=Z^{-1}(\beta,\delta h)\exp\left(-\beta\hat{H} + \beta \delta h \sum_{x=-L/2}^{L/2-1}\, x\, \hat{q}_{x}\right),
\end{equation}
with $Z(\beta,\delta h)=\text{Tr}\,\hat{\varrho}(\beta, \delta h)$.
Our aim is to compute the linear-response DC conductivity $\sigma(\beta)$, which we define
as the induced current density $\hat{j}^{(q)}_{0}(t)$ in the limit of vanishing bias $\delta h$,
\begin{equation}
\label{cond}
\sigma^{(q)}(\beta)=\lim_{t\to\infty}\lim_{L\to\infty}\lim_{\delta h \to 0}
\frac{1}{\delta h}\,\left\langle \hat{j}_0^{(q)}(t) \right\rangle_{\beta,{\delta h}}.
\end{equation}
Here the time propagation is governed by the Hamiltonian $\hat{H}=\sum_{x}\hat{h}_{x}$, and the expectation value of the current is
\begin{equation}
\label{curr}
\left\langle \hat{j}_{0}^{(q)}(t) \right\rangle_{\beta,{{\delta h}}} \equiv
\text{Tr}\left(\hat{j}_{0}^{(q)}(t) \hat{\varrho}(\beta,{\delta h})\right).
\end{equation}
By resorting to the Lieb-Robinson theorem, any local perturbation on a lattice with bounded finite-range interactions $\hat{h}$
propagates with a finite maximal velocity denoted by $ v_{\text{LR}}$. This permits to write Eq.~\eqref{cond} as a
single scaling limit $t\to\infty$, provided the system size is scaled in accordance with the Lieb-Robinson 
velocity $L=2v_{\rm LR}t$, yielding
\begin{equation}
\sigma^{(q)}(\beta)=\lim_{t\to\infty}\lim_{\delta h \to 0}
\frac{1}{\delta h}\left\langle \hat{j}_{0}^{(q)}(t)\right\rangle_{\beta,\delta h}.
\end{equation}
The average value of the current density $\left\langle \hat{j}_{0}(t)\right\rangle_{\beta,\delta h}$ (see Eq.~\eqref{curr})
can be written as a sum of averages $\left\langle \hat{j}_{0}(t)\right\rangle_{\beta,\delta h,q}$ over sectors
with a fixed value of the charge density $q$,
\begin{equation}
q=\frac{2\left\langle \hat{q} \right\rangle_{q}}{d-1} \in [-1,1].
\end{equation}
The DC conductivity $\sigma^{(q)}$ is accordingly decomposed as a discrete sum over the charge sectors,
\begin{equation}
\sigma^{(q)}(\beta)=\lim_{t\to\infty}\lim_{\delta h\to 0}\frac{1}{\delta h}
\sum_{q=-1}^{1} P(q,2 \,v_{\text{LR}} t)
\left\langle \hat{j}_0(t) \right\rangle_{\beta,\delta h,q},
\end{equation}
with a step size
\begin{equation}
\Delta q=\frac{2}{(d-1)L}.
\end{equation}
Here $P(q,L) $ denotes the unbiased probability of finding a state with charge density $q$ in a system of the length $L$.

For large times $t\to \infty$ we first expand the current in $q$-sector as~\cite{Ilievski2012}
\begin{equation}
\lim_{\delta h \to 0}\frac{1}{\delta h}\left\langle \hat{j}_0(t)\right\rangle_{\beta,\delta h,q}=
2\, \mathcal{D}^{(q)}(q) t+ {D}^{(q)}_v(q)+\mathcal{O}(t^{-1}),
\end{equation}
where $\mathcal{D}^{(q)}(q) $ is the finite-temperature Drude weight, and ${D}^{(q)}_{v}(q)$ denotes the leading
$\mathcal{O}(t^{0})$ sub-ballistic correction which is assumed to be positive,
cf. refs.~\cite{PhysRevLett.119.080602,Spohn_JMP}. This term will be subsequently disregarded.
Furthermore, in the high-temperature limit $\beta \to 0$, the probability factor $P(q,2 v_{LR} t)$ can be approximated by with
the Gaussian distribution by neglecting the contributions from the sectors which become suppressed
in the large-$t$ limit \cite{PhysRevLett.119.080602}
\begin{equation}
P(q,L)\approx \sqrt{\frac{6}{(d^2-1)\pi L}}\exp\left(-\frac{3(d-1)}{2(d+1)}\, q^{2} L \right).
\end{equation}
Replacing the system size $L$ with the Lieb-Robinson cone yields
\begin{equation}
\label{sum1}
\sigma^{(q)} \geq \lim_{t\to\infty} \sum_{q} \sqrt{\frac{3}{(d^2-1)\pi v_{LR} t}}\exp\left(-\frac{3(d-1) v_{LR}t}{(d+1)}\, q^{2} \right)\mathcal{D}^{(q)}(q)\, t.
\end{equation}
Taking the $ t\to \infty$ limit and converting the sum in Eq.~\eqref{sum1} to an integral, we find
\begin{equation}
\sigma^{(q)}\geq\frac{1}{6\, v_{\rm LR}}\frac{d+1}{d-1}\partial_{q}^2\,\mathcal{D}^{(q)}(q).
\end{equation}
Using Einstein relation $D^{(q)}=\sigma^{(q)}/\chi$, where $\chi$ denotes the static spin susceptibility, we obtain
the following lower bound
\begin{equation}
D^{(q)}\geq \frac{2}{\beta (d-1)^2\, v_{\rm LR}}\partial_{q}^{2} \mathcal{D}^{(q)}(q).
\end{equation}
In the infinite temperature limit, the scaled static susceptibility $\tilde{\chi}=\lim_{\beta \to 0}(\chi/\beta)$ reads
\begin{equation}
\tilde{\chi}=\frac{1}{12}\left(d^2-1\right).
\end{equation}

The derivative with respect to the chemical potential $h$ can be expressed as
\begin{equation}
D^{(q)}\geq\left(\frac{\partial q}{\partial h}\right)^{-2}\frac{2}{\beta (d-1)^{2}\, v_{\rm LR}}
\partial_{h}^2 \mathcal{D}^{(q)}(h)=
\frac{18}{\beta (d^{2}-1)^{2} \,v_{\rm LR}}\partial_{h}^{2} \mathcal{D}^{(q)}(h),
\end{equation}
where we have taken into account the relation
\begin{equation}
q=\frac{1}{d-1}\left(d \coth{(d\,h)}-\coth{(h)}\right).
\end{equation}

The logic of the above derivation generalizes to finite temperatures by taking into account a temperature-dependent
Gaussian approximation of the probability distribution $P(q,L) $ in the vicinity of the half filling,
\begin{equation}
\label{avg}
\frac{\sum_{x} \exp(-\beta E_{x,h})}{\sum_{x,h'} \exp(-\beta E_{x,h'})}\approx \exp\left(-f(\beta) h^{2}\,L\right).
\end{equation}
The finite temperature bound thus reads
\begin{equation}
D^{(q)}(\beta)\geq \frac{\partial_{h}^{2} \mathcal{D}^{(q)}(\beta,h)\Big|_{h=0}}{4 \chi(\beta) f(\beta) v_{\rm LR}}.
\end{equation}

Finally, we established the connection between the static susceptibility $\chi(\beta)$ and function $f(\beta)$.
In order to achieve this, we need to related the average \eqref{avg} with the grand-canonical average with respect to
$\hat{\varrho}(\beta, h)\simeq \exp(-\beta \hat{H}+2h\,\hat{Q})$. In the Gaussian approximation we have
\begin{equation}
\frac{\text{Tr}(\hat{\varrho}(\beta,h)) }{\text{Tr}(\hat{\varrho}(\beta,0))}\approx
\exp\left(-\frac{\beta}{2} \chi(\beta)(2h)^{2} L\right),
\end{equation}
implying
\begin{equation}
\label{fchi}
f(\beta)=2\beta \chi(\beta).
\end{equation}

\end{document}